\documentclass[manuscript,screen]{acmart}

\newif\ifdraft
\drafttrue 

\usepackage{adjustbox}
\usepackage{booktabs}  
\usepackage{adjustbox} 
\usepackage{xspace}
\usepackage{comment}
\usepackage{multirow}
\usepackage{amsmath}
\usepackage{amsfonts}
\usepackage{subfig}
\usepackage[font={bf,it,small},labelsep=colon,justification=centering]{caption}
\usepackage[super]{nth}
\usepackage{xurl}
\usepackage{setspace}
\usepackage{titlesec}
\usepackage[normalem]{ulem}
\usepackage{algorithm}
\usepackage[noend]{algpseudocode}
\usepackage[hang,flushmargin]{footmisc} 
\usepackage{xcolor}
\usepackage{todonotes}

\newcommand{\josh}[1]{\ifdraft{\textcolor{red}{[Josh: #1]}}\fi}
\newcommand{\saurabh}[1]{\ifdraft{\textcolor{blue}{[Saurabh: #1]}}\fi}

\newcommand{\dimitris}[1]{\ifdraft{\textcolor{orange}{[Dimitris: #1]}}\fi}

\newcommand{\arash}[1]{\ifdraft{\textcolor{teal}{[Arash: #1]}}\fi}

\definecolor{darkgreen}{rgb}{0.0, 0.2, 0.13}

\newcommand{\chop}[1]{}

\newcommand{\ie}{{\em i.e.},\xspace}

\usepackage{pifont}
\usepackage{multirow}
\usepackage{float}
\usepackage{subfig}
\usepackage{soul}

\AtBeginDocument{%
  \providecommand\BibTeX{{%
    \normalfont B\kern-0.5em{\scshape i\kern-0.25em b}\kern-0.8em\TeX}}}

\setcopyright{acmcopyright}
\acmJournal{CSUR}
\acmDOI{10.1145/3724113}

\begin{document}

\title{The Federation Strikes Back: A Survey of Federated Learning Privacy Attacks, Defenses, Applications, and Policy Landscape} 



\author{Joshua C. Zhao*}
\affiliation{%
  \institution{Purdue University}
  \city{West Lafayette, IN}
  \country{United States of America}
  }
  \thanks{*Other than the first two authors, all other authors are listed alphabetically. Corresponding author: Saurabh Bagchi.}
\email{zhao1207@purdue.edu}

\author{Saurabh Bagchi*}
\affiliation{%
  \institution{Purdue University and KeyByte}
  \city{West Lafayette, IN}
  \country{United States of America}}
\email{sbagchi@purdue.edu}

\author{Salman Avestimehr}
\affiliation{%
  \institution{University of Southern California and FedML}
  \city{Los Angeles, CA}
  \country{United States of America}}
\email{avestime@usc.edu}

\author{Kevin S. Chan}
\affiliation{%
  \institution{DEVCOM Army Research Laboratory}
  \city{Adelphi, MD}
  \country{United States of America}}
\email{kevin.s.chan.civ@army.mil}

\author{Somali Chaterji}
\affiliation{%
  \institution{Purdue University and KeyByte}
  \city{West Lafayette, IN}
  \country{United States of America}}
\email{schaterji@purdue.edu}

\author{Dimitrios Dimitriadis}
\affiliation{%
  \institution{Amazon}
  \city{Bellevue, WA}
  \country{United States of America}}
\email{dbdim@amazon.com}

\author{Jiacheng Li}
\affiliation{%
  \institution{Purdue University}
  \city{West Lafayette, IN}
  \country{United States of America}}
\email{li2829@purdue.edu}

\author{Ninghui Li}
\affiliation{%
  \institution{Purdue University}
  \city{West Lafayette, IN}
  \country{United States of America}}
\email{ninghui@purdue.edu}

\author{Arash Nourian}
\affiliation{%
  \institution{Amazon}
  \city{Sunnyvale, CA}
  \country{United States of America}}
\email{nourian@gmail.com}

\author{Holger R. Roth}
\affiliation{%
  \institution{NVIDIA}
  \city{Bethesda, MD}
  \country{United States of America}}
\email{hroth@nvidia.com}

\renewcommand{\shortauthors}{Zhao, et al.}

\begin{abstract}
  Deep learning has shown incredible potential across a wide array of tasks, and accompanied by this growth has been an insatiable appetite for data. However, a large amount of data needed for enabling deep learning is stored on personal devices, and recent concerns on privacy have further highlighted challenges for accessing such data. As a result, federated learning (FL) has emerged as an important privacy-preserving technology that enables collaborative training of machine learning models without the need to send the raw, potentially sensitive, data to a central server. However, the fundamental premise that sending model updates to a server is privacy-preserving only holds if the updates cannot be "reverse engineered" to infer information about the private training data. It has been shown under a wide variety of settings that this privacy premise does {\em not} hold.
  
  In this survey paper, we provide a comprehensive literature review of the different privacy attacks and defense methods in FL. We identify the current limitations of these attacks and highlight the settings in which the privacy of an FL client can be broken. 
  We further dissect some of the successful industry applications of FL and draw lessons for future successful adoption. We survey the emerging landscape of privacy regulation for FL and conclude with future directions for taking FL toward the cherished goal of generating accurate models while preserving the privacy of the data from its participants.

\end{abstract}




\received{6 May 2024}
\received[revised]{24 February 2025}
\received[accepted]{4 March 2025}

\maketitle

\section{Introduction}

Federated Learning (FL) is a popular learning paradigm that allows one to learn a Machine Learning (ML) model collaboratively. The classical structure of FL is with multiple clients each having their own local data, which they would possibly like to keep private, and there is a server that is responsible for learning a global ML model.
\footnote{The phrase ``The Federation Strikes Back'' is a playful nod to the \emph{The Empire Strikes Back}, symbolizing the ongoing arms race between privacy attackers and defenders in the federated learning domain, where data confidentiality stands at the heart of the conflict.}

One of the two primary reasons for the popularity of FL is that clients can keep their data private and still benefit from the combined learning across all of their data. (A second reason is the ``power of crowds'', i.e., many weak devices can come together to learn complex models, which would be beyond the compute power of any one client to learn on its own.) However, by 2018, the privacy of FL has been put into question questioned. In a set of seminal papers, it was shown that a central aggregator with access to gradients sent by the clients (which is standard in most versions of FL), then the aggregator can learn various things from these clients, each of which would be taken to effectively break the privacy of client data.~\cite{zhu2019deep,zhao2020idlg,melis2019exploiting}
The simplest form of this attack is where the aggregator can reconstruct the data of the clients from the gradient updates, to different degrees of fidelity. The attacker, the central aggregator here, has the unique advantage that it can observe individual updates from the clients over time and can control the view of the participants of the global parameters. Also worryingly, the attacker can even be one of the clients who can observe the global parameter updates, and can control his parameter uploads~\cite{hitaj2017deep}.

In this article, we survey the constant back and forth that has been roughly going on for the last 5 years in protecting the privacy of data through FL. We will have as one of the primary focus questions: what are the research challenges that need to be solved to achieve ``good enough'' protection for different application use cases. Our article takes a distinctive viewpoint of grounding the technical discussion of the attacks and the defenses in application use cases. For example, the kind of privacy expectation, and regulation, in the healthcare sector is very different from that in the finance or insurance sector, which again is distinct from that in Internet of Things (IoT). 

We provide a categorization of the attacks against privacy in FL (Section~\ref{sec:attacks}) and also a categorization of the defenses (Section~\ref{sec:defenses}). A coupled dimension is the kind of FL that is happening --- cross-device vs. cross-silo, horizontal FL vs. vertical FL, or derivatives of classical FL, such as, peer-to-peer learning or hierarchical FL. The discussion of attacks and defenses is meaningful only when we provide the context of what the threat model is, \ie what are the capabilities available to the adversaries and to the defenders, what aspect of the learning process is attacked (model, data, or something else), and what is the proportion of benign vs. adversarial entities. We provide this kind of foundation-setting discussion in Section~\ref{sec:problem-setting}. 

\noindent{\em Policy considerations.}
Our article goes beyond the technology landscape and also surveys the policy landscape (Section~\ref{sec:policy}), which is still developing. A central consideration of ML regulation, namely, regulation of the data, is central to FL since here we deal with data that is local, and possibly sensitive, to individual clients. Regulation of FL is therefore important and a start needs to be made soon. We hope that this discussion of regulation, in the context of the technical achievements to date and the expected deliverables, will lead to better regulation in the space of privacy in FL. 

\noindent{\em Privacy pushes-and-pulls.}
Given that being a privacy-preserving technology was a core reason for its introduction, it is vital that FL is indeed able to preserve the privacy of users. However, the process of sending gradients or model updates to a server has been shown to be less than secure. Recent works in membership inference, property inference, and even data reconstruction have demonstrated that vanilla FL is vulnerable to a multitude of privacy attacks. Defenses such as secure aggregation, differential privacy, or homomorphic encryption have also continued to be developed to prevent these attacks. Ultimately, the privacy expectation of FL is that an attacker, whether it is the server, an outsider, or a participating client, is not able to infer any private information about the local training data of any of the clients. In this work, we discuss current privacy attacks and pinpoint the vulnerabilities that make them possible. We then highlight the defenses that can prevent the attacks or make them more difficult along with the limitations and drawbacks of each.


\section{Problem Context}
\label{sec:problem-setting}
\subsection{Federated learning process}
FL has emerged as a decentralized training framework that addresses the dual challenges of leveraging computational resources at the network's edge and preserving data privacy. Central to FL is its unique approach to collaborative model training, which aggregates model updates from participating clients rather than transmitting the raw data of the clients. This method not only leverages the vast computational power distributed across edge devices but also aligns with the increasing demands for data privacy amidst the exponential growth and insatiable appetite for data of modern AI systems.
At the heart of FL is the distributed training of machine learning models, where the crux lies in the \textit{exchange of model updates} with a central server, sidestepping the need to share sensitive training data. 

The vanilla FL training process typically involves the coordinated interplay of two principal actors: the central server and the participating clients. The central server orchestrates the workflow by selecting the clients, dispatching the global model to the participating clients for local refinement, and then aggregating the updates before finally updating the global model again. Conversely, clients engage in a straightforward yet crucial role. They receive the global model from the server, apply training updates using their local datasets, and then transmit their updates back to the server. 

For an FL training cycle, the operations unfold as follows:
\begin{enumerate}
    \item \textbf{Client Selection and Model Dispatch}: The central server selects a subset of $k$ out of $N$ available clients, dispatching the current global model parameters, $G_{t}$, to them. The selection count varies based on the FL scenario:
    \begin{itemize}
    \item In \textit{cross-device FL}, typically involving lightweight devices like smartphones or IoT devices, the selection ranges from $100$s to $1000$s. Device communication and computation are typically more unreliable.
    \item In \textit{cross-silo FL}, involving entities like data centers or hospitals, a tighter group of $5$--$10$ clients is chosen, reflecting the higher trust, collaboration levels, and substantial computational resources of these participants.
    \end{itemize}
    
    \item \textbf{Local Training and Update Generation}: Participating clients refine the global model $G_{t}$ using their local data across one or more iterations. This process yields updates, either as computed gradients, $\nabla G_{t,i}$, or direct model parameter modifications, which are then sent back to the central server.
    
    \item \textbf{Aggregation and Global Model Update}: The server aggregates these client updates using methods such as a weighted mean or other aggregation method to form the new global model parameters, $G_{t+1}=G_{t} + \sum_{i=1}^{k} (w_i \cdot \nabla G_{t,i})$. This prepares the model for the next distribution, refinement, and aggregation cycle.
    
    \item This iterative process repeats until satisfactory model performance is reached, i.e., convergence is reached to a reasonable degree.
\end{enumerate}

\subsection {Enhancing Federated Learning: Key Considerations for Operational Integrity and Security.}
Several additional considerations come into play, ensuring the smooth and secure operation of this collaborative model training process. Here are some of the important considerations.\\
\textbf{Client eligibility and selection:}
One of the initial steps in the FL cycle involves the selection of participating clients. In scenarios categorized under cross-device FL, where the participants are typically mobile or IoT devices, certain prerequisites must be met before a device is deemed eligible. These prerequisites often include having sufficient battery life to complete the training tasks without interruption and a stable network connection to facilitate seamless communication with the central server. These criteria are essential to prevent dropout or delays in the training process.\\
\textbf{Security enhancement through encryption:}
Once selected and during the model refinement phase, enhancing security becomes paramount. To this end, clients may employ encryption techniques to the updates they generate. This encryption ensures that the gradients or model parameter modifications are securely transmitted to the central server, safeguarding the data against eavesdropping or tampering attempts. This layer of security is crucial in preserving the integrity of the data and the privacy of the clients' information.\\
\textbf{Robust aggregation by the central server:}
The central server plays a critical role in aggregating the updates from all participating clients. Given the potential for compromised devices or malicious actors within the network, the server might implement robust aggregation techniques~\cite{sharma2021tesseractgradientflipscore, sharma2023flair}. These techniques are designed to identify and neutralize the influence of any anomalous updates that could skew the model's learning in an undesirable direction. By employing such strategies, the server ensures that the aggregated update accurately reflects the collective learning from all legitimate participants, thereby reinforcing the reliability and effectiveness of the process.

\subsection{Variants of Federated Learning}
\subsubsection{Cross-device and Cross-silo Federated Learning:} FL can be broadly categorized into cross-device and cross-silo contexts, each presenting unique scalability and deployment challenges. 
Cross-device FL typically operates with a vast network of personal devices, such as smartphones, wearables, or IoT devices, each participating in the learning process. This variant is marked by its massive scale, involving potentially \textit{millions of devices}, each contributing data for model training. The finance sector, for example, could leverage FL to enhance fraud detection systems by learning from transactions across countless mobile banking apps without centralizing sensitive financial data. However, cross-device FL still has many challenges, primarily due to the limited computational power and intermittent connectivity of devices, not to mention the heightened concerns over data privacy and security.

Conversely, cross-silo FL operates on a smaller scale with fewer, resource-ample entities like hospitals, featuring stable communications and higher computational resources. In healthcare, for instance, cross-silo FL enables hospitals to collaboratively improve diagnostic models by learning from diverse patient datasets while ensuring that sensitive medical records remain within hospital premises. Similarly, in finance, banks can utilize cross-silo FL to jointly develop more accurate credit scoring models without exposing their clients' financial details to competitors. This form of FL, while offering more control and resources, still necessitates careful navigation of regulatory, privacy, and interoperability challenges. We discuss the applications of various privacy-preserving defenses for each category in-depth in Section~\ref{sec:defenses}.
\subsubsection{Federated transfer learning:}
Federated Transfer Learning (FTL) extends FL by facilitating knowledge sharing across different domains or tasks, aiming to enhance model performance where direct data sharing is infeasible. Unlike FL, which focuses on aggregating updates to refine a global model, FTL emphasizes transferring knowledge from a well-resourced source domain --- rich in labeled data --- to a target domain with limited labeled resources. This cross-domain learning is achieved by sharing model parameters or learned features instead of raw data, thus inherently supporting privacy by keeping the data localized while only model information traverses domains.



FTL is particularly useful in scenarios where data cannot be pooled together due to privacy concerns but where tasks across domains are related. For example, in healthcare, models trained on patient data from one hospital on one task can be adapted to improve predictions in another hospital on a different task, despite differences in patient demographics or data collection practices. Similarly, in finance, the knowledge from credit scoring models in one region can be transferred to enhance models in another region.

\subsubsection{Hierarchical Federated Learning}

Hierarchical Federated Learning (HFL)~\cite{abad2020hierarchical,feng2021min,liu2020client,briggs2020federated,liu2022hierarchical} introduces a multi-tiered aggregation framework, significantly enhancing the scalability and efficiency of collaborative learning models. This system incorporates several levels of servers, from edge parameter servers to a final cloud parameter server, facilitating a more structured and efficient communication and aggregation pathway. By leveraging edge servers for initial aggregation~\cite{liu2020client}, HFL mitigates the communication bottlenecks often encountered in traditional FL, enabling faster, more energy-efficient training processes. In response to the limitations observed in both FL and P2PL --- where FL's centralized model can pose privacy risks, and P2PL's decentralized nature complicates integrity and consensus --- HFL emerges as a powerful hybrid solution. It employs a hierarchical structure in which clients communicate updates to edge servers which then aggregate these updates before further communication among themselves or with a higher-level server. This architecture not only reduces the privacy concerns associated with a single central server but also addresses the communication and consensus challenges inherent in a fully decentralized setup.

\subsubsection{Horizontal and Vertical Federated Learning}

In FL, data can be partitioned in two primary ways: horizontally or vertically. This distinction significantly impacts the learning process and the approach to privacy and efficiency in collaborative learning environments. Horizontal Federated Learning involves clients that have data sharing the \emph{same feature space} but differ in the \emph{samples} or \emph{IDs}. This is common in scenarios where different entities collect similar types of data across various subjects.\\
\textbf{Example:} Consider multiple hospitals collecting health data. Each hospital has its own set of patient data, with the feature space (types of health data) being consistent, but the actual data (patient records) varying.
Mathematically, this can be represented as:
\begin{equation}
\text{Let } D_i \text{ and } D_j \text{ be datasets from hospitals } i \text{ and } j \text{ respectively, where}
\end{equation}
\begin{equation}
D_i, D_j \subseteq \mathcal{X} \times \mathcal{Y}, \quad \text{and} \quad \mathcal{X}_i = \mathcal{X}_j, \quad \text{but} \quad \mathcal{Y}_i \neq \mathcal{Y}_j
\end{equation}

\noindent Vertical Federated Learning involves clients that have data with the \emph{same IDs} but inhabit different \emph{feature spaces}. This setup is typical in collaborations between entities that collect different types of data about the same subjects.\\
\textbf{Example:} A bank and a hospital share data about the same individuals, with the bank holding financial information and the hospital holding medical records. They share IDs but have vastly different feature spaces.
\begin{equation}
\text{Let } D_a \text{ from a bank and } D_b \text{ from a hospital, where}
\end{equation}
\begin{equation}
D_a, D_b \subseteq \mathcal{X} \times \mathcal{Y}, \quad \text{and} \quad \mathcal{Y}_a = \mathcal{Y}_b, \quad \text{but} \quad \mathcal{X}_a \neq \mathcal{X}_b
\end{equation}

\subsubsection{Data Distribution Attributes and Differences}

\textbf{IID Data Distributions:} In scenarios where data is considered IID, each client's dataset is assumed to be drawn from the same probability distribution. This uniformity suggests that datasets across varied domains, such as those of a bank and a hospital, exhibit similar statistical properties and structures, despite their distinct contexts. \textbf{Non-IID Data Distributions:} On the other hand, non-IID data distributions reflect situations where data across clients originate from heterogeneous distributions. This diversity is rooted in demographic variances, regional characteristics, or the intrinsic nature of the collected data.

\subsubsection{FedSGD and FedAvg: Core FL algorithms} The two standard FL algorithms are FedSGD and FedAvg~\cite{mcmahan2017communication}, mainly differ in the number of local iterations. FedSGD, the federated adaptation of stochastic gradient descent, operates on the principle of minimalism. It mandates that clients engage in a singular iteration of training on a localized data batch before transmitting the resultant gradient to a central server. This server, in turn, aggregates the gradients from all participating clients, adjusting the weights in proportion to each client's contribution to the total dataset. 

Advancing beyond FedSGD, FedAvg introduces multiple local iterations and/or epochs at each client before communicating updates to the server. These updates can comprise either the computed gradients over all local iterations or the updated model parameters themselves. 
The algorithm structure inherently minimizes the frequency of communication with the server, but also helps with privacy concerns by limiting the server exposure to initial and final model states only. This characteristic is surprisingly helpful in contexts where data sensitivity is paramount, as it effectively conceals the intermediate computational states from potential attackers. 

\subsection{Technical Challenges for Federated Learning}
Federated learning expects computational and bandwidth resources to fully leverage the private data. Multiple federated learning algorithms have proposed solutions for  efficient communication and/or compute resource scarcity. However, most of the deployed platforms are not built to address such requirements, making them the wrong choice for seamless, scalable federated learning. These platforms are often limited in resources, competing with other onboard functionalities, making scheduling an additional challenge. Earlier solutions were performing federated learning computations when the devices were idle, during the night and plugged in a power outlet. However, this can create biased models, where some of the data distributions are largely ignored.

\subsubsection{Implications on Security of the Learning}
Federated Learning introduces Security risks mainly for two reasons: the private data cannot be audited or curated, enabling ``bad actors’’ to alter their data according to the end-goal, i.e. ``training data attacks’’. Also, the models are exposed during the communication process, introducing security risks while providing opportunities for backdoor attacks. These attacks aim at ``influencing’’ the model parameters to a particular behavior, or simply degrading the overall model performance.

\subsubsection{Data and Device Heterogeneity}
Data and device heterogeneity can lead to biased models. The private data distributions and the available computational resources can vary widely between different clients, which leads to issues such as global model drift or the need to adopt multiple model architectures. 
FedAvg often shows performance disadvantages to highly mismatched data distributions, and several approaches have been proposed to address this issue. 
Besides any mismatch between distributions, the lack of labels during training can also produce suboptimal training conditions. This problem can be either mitigated by Knowledge Distillation~\cite{ChJoDi23}, heuristics~\cite{WWSC24} or asking users to label their data~\cite{Augenstein+20}. A basic assumption of the early FL algorithms was the local model structure is uniform across all clients. However, clients can differ in their computational resources and data distributions.

\subsection{Threat model}
For privacy attacks within the domain of FL, the the threat landscape is complex and varies significantly based on the nature of the attack. The architecture of FL inherently poses unique challenges in safeguarding privacy due to the decentralized nature of data processing. 
Understanding the nuances of these threats is crucial for developing robust defense mechanisms. 
At the heart of FL, the central server embodies a critical point of vulnerability. It plays a dual role: distributing the learning model to clients and aggregating their updates. This centralized control positions the server as a potent target for attacks, making it the focal point for the strongest privacy threats in FL. 
This is split into two broad types of threats as follows:\\
\textbf{Central Server Threats:}
\textit{Honest-but-Curious Server:} This scenario presupposes a server that adheres strictly to the prescribed FL protocol without any malicious alterations. However, it harbors intentions to infer private information from the updates received from clients. Despite its non-intrusive facade, this model represents a significant privacy threat as the server possesses the capability to analyze aggregated data and potentially extract sensitive client information without overtly compromising the integrity of the training process.\\
\textit{Malicious Server:} A more daunting threat emerges with the malicious server, which actively seeks to undermine the privacy and integrity of the FL process. This adversary is not bound by the ethical constraints of the honest-but-curious model and may engage in manipulative tactics such as altering client model parameters, tweaking the model architecture, or distorting the training process. The objective is clear: to siphon off private data from clients or to inject biases into the model, thereby compromising both privacy and model fidelity.

\noindent \textbf{Client-Side Threats:}
A parallel threat model arises due to the clients. Corresponding to the server case, here also the clients can be:
\textit{Honest-but-Curious Clients:} Parallel to the server-side threat, clients under this model operate within the bounds of the FL protocol but with ulterior motives. They aim to intercept and analyze communications between other clients and the server, with the ultimate goal of reconstructing the local data of their peers. This form of eavesdropping introduces a subtle yet potent risk to data confidentiality within the FL ecosystem.\\
\textit{Malicious Clients:} Beyond mere curiosity, malicious clients actively engage in sabotaging the FL process. By submitting falsified updates, these adversaries attempt to derail the collective learning effort, leading to a degraded model utility. While this behavior does not directly infringe on privacy, it undermines the integrity and reliability of the FL system, posing a significant challenge to maintaining model quality.

\noindent \textbf{Complexities of Misbehaving Clients:}
The threat landscape is further complicated by the variability in the behavior of misbehaving clients. Two critical factors need consideration: the proportion of clients engaged in malicious activities and the extent of their collaboration. Misbehaving clients may operate independently, presenting a scattered threat. However, the danger amplifies when they collude, orchestrating their attacks to maximize disruption or data extraction. This coordinated effort can significantly exacerbate the challenges in defending against privacy breaches and ensuring the integrity of the FL process~\cite{sharma2023flair}.
%


%

\subsection{Metrics measuring privacy}

\textbf{Membership inference attacks - accuracy.} The goal of membership inference attacks is to infer whether a particular instance is used in the training of a target model. A balanced evaluation set, which consists of the same number of members and non-members, is used by the membership inference attack evaluations in most existing literature.
For membership inference attacks, the attack accuracy measures the attack effectiveness over the whole data population~\cite{shokri2017membership}. 
Attack AUC can also be used to measure the privacy leakage in terms of membership inference over the whole population. 
The higher the attack accuracy is, the more privacy leakage the target model has on its training set for the whole set.

\textbf{Membership inference attacks - TPR at a low FPR.} Accuracy and AUC measures the average privacy vulnerability across all instances.  In the context of privacy, one usually wants to ensure that every individual's privacy is protected.  It is thus necessary to measure membership leakage threat for the most vulnerable instances.  In~\cite{carlini2022membership}, it was proposed to use true positive rate (TPR) at a low false positive rate (FPR) to measure effects of MI attacks.  
Consider, for example, TPR at FPR$=0.001$.  If $TPR$ is also around $0.001$, then that means the adversary cannot do better than random guessing even when they are very confident about MI inference.  On the other hand, if If $TPR$ is around $0.01$, then an adversary would be able to correctly identify (a small percentage of) members with a $10:1$ odds ratio. 

\textbf{Data reconstruction attacks - reconstruction quality.} One important metric for gauging the success of data reconstruction attacks is in the similarity of the reconstructions to the ground truth images. This measure can be visual and is typically quantified through image similarity metrics such as PSNR (peak signal-to-noise ratio), SSIM (structural similarity index measure), or L-PIPS (learned perceptual image patch similarity). Image identifiability precision (IIP) was also introduced as a potential metric for reconstruction quality~\cite{yin2021see}. Instead of per-image calculatioins, compared to all images in the ground truth batch, IIP finds the nearest neighbor of each reconstructed image. The metric is quantified as the fraction of reconstructions that have the ground truth image as their nearest neighbor. Additionally, \cite{zhao2024leak} discusses the use of leaked data for training models in downstream prediction tasks related to the image similarity metrics, where even reconstructed images with the lowest metric scores (e.g., PSNR$<12$, SSIM$<0.2$) were still shown to improve the accuracy of trained models~\cite{zhao2024leak}.

\textbf{Data reconstruction attacks - leakage rate.} Another metric used to quantify success of data reconstruction attacks is leakage rate. While leakage rate can be simply defined as the number of leaked images, quantification can still vary. Some methods, such as linear layer leakage, can result in leaking an image consisting of multiple client images overlapping. Here, leakage rate is typically defined as the number of reconstructions with only have a single, non-overlapping image leaked. Although methods such as gradient inversion typically rely on image similarity to measure quality, leakage rate can also be used based on the number of images above a given threshold for any metric (e.g., PSNR$>16$)~\cite{hatamizadeh2023gradient}.

\section{Relation to Prior Surveys}
\label{sec:related}

Federated learning was originally introduced as a method to collaboratively train machine learning models while allowing users to keep their data local and private. Despite this, many prior works have shown that the updates sent to a server still inherently include sensitive information that can be used to leak either properties of the data or reconstruct them directly. Following this, much work has also been done in the area of FL privacy defense. To motivate our work, we first compare the prior surveys in FL privacy and discuss the missing pieces. Table~\ref{tab:related_surveys} shows a summary of previous surveys compared to this work.

Many surveys have looked at the privacy aspect of FL. The works of \cite{el2022differential,khan2023federated,li2020federated} discuss privacy defense in FL with little discussion on privacy attacks. Other surveys~\cite{mothukuri2021survey,zhang2022challenges,yin2021comprehensive,bouacida2021vulnerabilities,chen2022federated,sikandar2023detailed,liu2022threats} look at privacy attacks, but are missing discussion on more recent catastrophic attack methods such as linear layer leakage for data reconstruction or membership inference attacks that do not require training data. \citet{chen2023privacy} discusses the trade-off between privacy and fairness in FL. While the methods of optimization and closed-form linear layer leakage attacks are discussed, the work still misses other key data reconstruction methods such as multi-round disaggregation or attacks targeted against secure aggregation. Similarly, \citet{rigaki2023survey} discusses privacy attacks in machine learning but also lacks many current privacy attacks specific to FL. \citet{lyu2022privacy} is the closest survey to ours, focusing on security and privacy attacks and defenses in FL. However, important discussion on the key limitations of attacks and defenses is missing. Similarly, recent, powerful data reconstruction attacks such as~\cite{fowl2022robbing,wen2022fishing,lam2021gradient,kariyappa2022cocktail} are missing.

Our work bridges this gap on the recent and rapid development of privacy attacks and defenses and provides a comprehensive discussion on current privacy threats (data reconstruction, membership-inference, and property inference attacks) and defenses (differential privacy, secure aggregation, homomorphic encryption, and trusted execution environments) while highlighting the limitations of these methods along with areas where improvement is needed. Additionally, to the best of our knowledge, we are the only work that discusses current industry applications utilizing FL in a variety of industry sectors including healthcare, finance, and IoT edge applications. Similarly, we are the only work to include crucial discussion on government data privacy policies and their relation to FL. We close our work by looking ahead and hypothesizing the key developments that would be necessary for FL to become a privacy-preserving technology that addresses user privacy concerns and policy mandates in real-world deployments.


\begin{table}[]
\caption{\label{tab:related_surveys} Comparison of federated learning privacy surveys}
\begin{tabular}{lcccc}
\hline
\multicolumn{1}{c}{\multirow{2}{*}{\textbf{References}}} & \multicolumn{2}{c}{\textbf{Privacy}} & \multicolumn{2}{c}{\textbf{Industry}} \\ \cline{2-5} 
\multicolumn{1}{c}{}                            & {Defenses}      & {Attacks}     & {Applications}     & {Policy}    \\ \hline
                                   Li et al~\cite{li2020federated}             &     \ding{55}         &              &                  &           \\
                                   Ouadrhiri et al~\cite{el2022differential}             &     \ding{55}         &              &                  &           \\
                                   Chen et al~\cite{chen2023privacy}             &     \ding{55}         &  \ding{55}            &                  &           \\
                                   Lyu et al~\cite{lyu2022privacy}             &     \ding{55}         &   \ding{55}           &                  &           \\
                                   Rigaki et al~\cite{rigaki2023survey}             &     \ding{55}         &   \ding{55}           &                  &           \\
                                   Our work             &  \ding{55}            &     \ding{55}         &       \ding{55}           &    \ding{55}       \\ \hline
\end{tabular}
\vspace*{-3mm}
\end{table}

\section{Application Use Cases and Policy Drivers}
\label{sec:use-cases}

Federated learning has multiple applications across various industries, including healthcare, finance, automotive, and IoT/Edge devices. In healthcare, FL can be used to train models on patients’ data without compromising privacy, especially when using differential privacy. Similarly, in the financial industry, we can employ FL to train fraud detection models without exposing sensitive financial data to potentially malicious entities. In the automotive industry, FL can be employed to train autonomous driving models on real-life data while avoiding the costs of transferring huge volumes of data back to the servers. Table~\ref{tab:applications_cross} further summarizes the application domains along with the commonly applicable setting (cross-device and cross-silo) for deployment.

FL has been successfully deployed in industry settings~\cite{Ge2021FailurePI}~\cite{Imperial}~\cite{9529467}~\cite{Sun_Tang_Yang_Yu_Wang_Ding_Li_Yu_2024}. One example of successful adoption comes from predictive maintenance and failure prediction, which are crucial for large-scale production lines.  They require real-time failure monitoring and updates from multiple resources across an enterprise that are often guarded by different privacy and security policies. 
Bosch~\cite{Ge2021FailurePI} has used many federated algorithms in various horizontal FL and vertical FL production scenarios such as predictive maintenance. Through this, Bosch was able to improve the life expectancy of production line machinery over time without significant overhead compared to centralized learning approaches.
Other companies~\cite{Imperial} have also used FL in similar use cases to drastically reduce the cost of halting in the production line due to unexpected failures. 


\begin{table}[ht]
\caption{\label{tab:applications_cross} Characterization of federated learning application domains into cross-device and cross silo deployments}
\begin{tabular}{lcc}
\hline
\textbf{Application Domain} & \textbf{Cross-Device} & \textbf{Cross Silo} \\ 
\hline
Healthcare &     \checkmark & \checkmark \\
Finance &              &  \checkmark \\
IoT / Edge &     \checkmark         &  \checkmark \\
Autonomous Driving &     \checkmark         &    \\ \hline
\end{tabular}
\vspace*{-3mm}
\end{table}


\subsection{Application of FL in Healthcare}
\label{subsec:healthcare}

Over the past few years, the automation of the healthcare industry has surged, driving an increasing demand for advanced ML models.  Healthcare professionals are leveraging technology to enhance patient care and better meet emerging global demands while managing vast and sensitive datasets. Training effective healthcare models, however, requires access to diverse data sources, and sharing such information is increasingly challenging under strict privacy regulations. FL offers a promising solution by enabling collaborative model training across institutions without exposing sensitive local data. With FL, participating organizations can develop a global model based on their in-house data, leading to more accurate and diverse ML models~\cite{eu2019trustworthy} while complying with regulatory frameworks and mitigating data silo issues.
%

Several real-world efforts have been made around FL in the healthcare sector, including proof-of-concept projects and the establishment of long-running consortia~\cite{rieke2020future,rehman2023federated}.
For example, in the United Kingdom, the FLIP project empowers AI researchers to develop clinical applications using NHS patient data without transferring the information outside the hospital network, thus addressing the challenge of confidentiality in healthcare~\cite{FLIP}. 
%
In France, the HealthChain consortium~\cite{healthchain} aims to address the bottleneck in accessing medical data for AI models that could lead to medical breakthroughs. 
In Germany, the Trustworthy Federated Data Analytics (TFDA) project~\cite{TFDA} aims to address the challenges posed by centralized data structures in the context of growing need for more data for machine learning. Recognizing the drawbacks and threats associated with data centralization, TFDA focuses on implementing decentralized cooperative data analytics architectures, specifically within and beyond the Helmholtz research community. 

The MELLODDY project~\cite{heyndrickx2023melloddy,oldenhof2023industry}, co-funded by the European Union and EFPIA companies, has successfully demonstrated the feasibility of large-scale collaborative artificial intelligence (AI) for drug discovery. By leveraging a large collection of small molecules, the project utilized federated and privacy-preserving machine learning to achieve more accurate predictive modeling, leading to potential efficiencies in drug discovery. 
The German Cancer Consortium’s Joint Imaging Platform (JIP)~\cite{scherer2020joint} is a strategic initiative within the German Cancer Consortium (DKTK) with the goal of establishing a technical infrastructure to facilitate modern and distributed imaging research. Emphasizing the use of contemporary machine learning methods in medical image processing, the project aims to enhance collaboration among clinical sites and support multicenter trials. 
Internationally, the Federated Tumor Segmentation (FeTS) initiative~\cite{fets,pati2021federated} is developing one of the largest federations of healthcare institutions. This initiative aims to gain knowledge for tumor boundary detection from diverse patient populations without sharing patient data. 

In a collaborative effort involving seven clinical institutions worldwide, an early study explored using FL to construct robust medical imaging classification models for breast density based on BI-RADS~\cite{roth2020federated}. Despite significant differences in datasets across sites, including mammography systems, class distribution, and data set size, the results demonstrated successful AI model training in a federated manner. 
The EXAM project~\cite{dayan2021federated} aimed to build a common predictive model to estimate the oxygen treatment requirements of patients arriving at the emergency department (ED) with symptoms of COVID-19 for the next 24h and 72h periods. A consortium of 20 hospital sites across the globe participated in this real-world application of FL. 
Furthermore, institutions such as The American College of Radiology (ACR) enable privacy-preserving AI utilizing FL technology~\cite{acr}.




The concept of federation has deep roots in computational genomics~\cite{chaterji2019federation}. Early work on federating computational resources---such as CPUs, GPUs, and FPGAs---and datastores demonstrated that distributed processing within unified portals could simplify data management and minimize extensive data transfers. This pioneering research remains relevant today by enabling the efficient handling of large-scale genomic data while ensuring privacy and security, paving the way for advanced federated approaches in healthcare. In particular, the seminal work on federated compute infrastructures for genomics highlights challenges such as heterogeneous data formats and disparate computational environments. By standardizing data exchange protocols and developing interoperable interfaces, early federated systems achieved seamless integration of diverse resources and minimized the need for large-scale data transfers. These innovations not only enhanced data security through localized processing and robust access controls but also laid the groundwork for overcoming technical and regulatory barriers—a legacy that informs and strengthens current FL applications in healthcare. Furthermore, computer systems research toward optimizing databases such as SOPHIA~\cite{sophia2019} has leveraged these federated principles to tackle dynamic workload challenges in real-world settings. The SOPHIA framework optimizes the reconfiguration of clustered NoSQL databases for applications like MG-RAST---a global-scale metagenomics repository---thereby ensuring high throughput and data availability despite fluctuating workloads. This integration of federated techniques with adaptive database management demonstrates how cross-silo federation can enhance scalability, performance, and resilience in modern database systems, making it a critical enabler for both computational genomics and healthcare applications.

An example of cross-device federated learning in the healthcare setting is action recognition performed on heterogeneous embedded devices~\cite{jain2021federated}. In healthcare, such methods can be applied to patient monitoring tasks, including fall detection or tracking physical activity during rehabilitation. Each device processes its local sensor or video data and contributes to a collaboratively trained global model without sharing raw data. This approach effectively addresses the challenges posed by devices with varying computational capabilities. In particular, asynchronous federated optimization allows each device to send its model updates to the server as soon as its local training is complete, rather than waiting for a synchronized round. This reduces idle time and mitigates the straggler problem---where slower devices hold back the overall training progress---thus accelerating the training process while maintaining robust model convergence.

\subsection{Application of FL in Finance}
\label{subsec:fintech}

The data privacy and protection laws are becoming ever more restrictive, especially in industries like the Financial sector. This allows consumers and businesses to trust each other to keep the data safe and secure. With traditional ML, businesses dependent on FinTech face several issues, such as getting clearance and lawful consent,  the governance of the customers' data, and finally, the cost in collecting and of managing such data. FL provides an efficient way for managing this data by keeping it local, on the financial institutions servers. FL is an encrypted and distributed machine learning approach that allows joint training on decentralized data where participants do not need to share it. 

One of the most promising applications in financial services is anti-money laundering (AML). In a typical money laundering scheme, criminals try to conceal the origin, identity, and destination of  financial transactions to look normal as they are from a legitimate source.  AML solutions allow financial institutions  to prevent, detect, investigate and report any activity that is suspicious of  money laundering and stop criminals from illegally obtained funds as legitimate income. The estimated amount of money laundered globally in one year is 2 - 5\% of global GDP (\$800 billion - \$2 trillion)~\cite{UN-AML}. Failure to prevent  money laundering can endanger the integrity and stability of global financial system. Financial institutions are required to put strict policies and measures to prevent, detect, and report laundered money and combat these crimes. 
However, each financial institution typically only has access to its own transnational data and can not prevent and detect complex money laundry cases where money routed through multiple financial institutions could look legitimate to individually involved institutions in the chain.  
Cross-silo Federated AML can greatly help financial institutions to create better predictive models with broader context for identifying bad actors. Such models learn more global and deeper correlation identifiers for bad actors and money laundering specific actions based on contributed data provided by a consortium of financial institutions.

Another application of FL is in insurance underwriting. The main use cases of FL in insurance include origination- predicting if a person contracts an insurance offer, insurer’s risk assessment, fraud detection, and claim processing. Due to soiled data in different companies on different demographics or customers movement across insurance companies, insurers can significantly improve their predictive models for these use cases by utilizing FL.  Claim history processing is of the most challenging use cases that can lead to more accurate and efficient operations. Insurance companies often need to have access to previous claims history of an insurer in details to give a better pricing or benefits to new clients. However, having access to such historical data in a privacy preserving manner is challenging due to regulations governing the insurance industry. Insurance companies can leverage cross-silo FL to privately train their models without compromising customers privacy and satisfying the governing regulations. Such models lead to more accurate offers while helping customers get the benefits of having a long-term history of free claims across different categories.  

\subsection{Application of FL in IoT or at the Edge}
\label{subsec:iot}

The Internet of Things (IoT) is the current concept of ubiquitous internet connectivity of all devices (refrigerators, toasters, doorbells, street cameras, ..., the list is unending). As the data from these devices scan come from anywhere, the privacy and security of these devices and the data is paramount. However, there is strong interest from companies and individuals to exploit the available data for constructive applications to benefit more than just the immediate use case and users. One prominent example is the Google keyboard (Gboard) prediction function \cite{hard2018gboard}, where Google improves the performance of their mobile keyboard prediction by gaining information from user interaction. This example highlights information privacy in that improvements to this technology will benefit all users, although users are only willing to contirbute if their privacy is ensured. Therefore, FL techniques that exploit data and maintain user and data privacy are attractive for IoT applications \cite{zhang2021fliot}.

In such scenarios and related domains, it is well understood that communication and computing resources are two of the bottlenecks in real-life FL applications. These devices are often used for tasks other than FL, so the FL task usually competes with these other ones for computing resources, making scheduling and coordinating even more complicated. Asynchronous strategies like PAPAYA~\cite{huba2022papaya} appear to address such issues. Other scenarios include heterogeneity in computing resources \cite{diao2021heterofl} or drift in local device clocks \cite{koo2009tale}, leading to asynchrony caused by varying compute and resulting in delays in the aggregation step. Methods to account for or dynamically adapt to changing available resources will greatly enhance the utility of FL techniques in these scenarios. The increased need for bandwidth for FL iterations has led to significant research efforts in reducing it. Methods combining optimizers such as FedAvg with sparsification and/or quantization of model updates to a small number of bits have demonstrated significant reductions in communication overhead. However, it still remains an open research question on how communication overheads can be further reduced and whether any of these methods or their combinations can come close to providing optimal trade-offs between communication and accuracy. For example, the Fed-ET approach communicates using smaller models \cite{Cho+22}. Another class of approach to maximizing bandwidth usage in FL approaches include device selection \cite{perazzone2022fl}. Although efforts in these situations are motivated by increasing communications efficiency, device selection methods can potentially provide device privacy in terms of revealing which nodes are contributing to the FL process. Moreover, there has also been recent work that allows control of the information exchange to limit communications, but can similarly be used to provide some privacy of node utility to FL algorithms \cite{wang2023flcontrol}.

Additionally, IoT has been promoted as a use-case for edge networks where bandwidth, storage, and compute are limited. These systems operate in a variety of environments often characterized a stringent operational requirements, despite having to cope with sophisticated adversarial influences and constraints in both computational and network resources. These environments include resource-constraints inhibiting information exchange to other nodes, specifically a central node with which centralized computation and decisions can be conducted. Federated approaches have been proposed for use in sensor networks, swarms of unmanned aerial vehicles and on coalition networks for shared situational awareness. They have also been proposed for FL at the edge by reducing the network overhead incurred by such learning processes \cite{wang2019fledge}. The promise of IoT and edge computing will continue to be revealed, with FL techniques potentially being a main driver of the technologies.

\section{Privacy-Related Policies}
\label{sec:policy}

Most of the existing regulations on safeguarding consumer privacy, both nationally and internationally, were conceived in a pre-AI/ML era. Currently, they have not been able to keep up with the rapid evolution of AI/Ml and their privacy implications. The General Data Protection Regulation ("GDPR")~\cite{GDPR} from the European Union governs data protection and privacy for all citizens within the EU and imposes strict obligations on data controllers and processors.  GDPR contains the \textit{principle of data minimization} in article 5.1 (c), which requires that data which is not necessary to achieve the intended purpose cannot be collected, stored, or otherwise processed lawfully. Since FL restricts the transfer of raw data to a central location, it provides more compliance with the principle of data minimization than other approaches that require the transfer of all raw data to a central location for model training. FL also protects the collected data from unwanted processing that is not compliant with the collection purpose, satisfying the \textit{principle of purpose limitation}, as described in article 5.1.(b). 

In the United States, the state of Privacy and AI laws is tracked by the EPIC project, which shows a recent large number of states that have effective privacy laws~\cite{EPIC}. Each of these privacy laws has specific requirements for accessing and processing personal data. Even FTC is enforcing the same privacy rights for companies using AI~\cite{FTC}.
At the federal level, sector-specific privacy polices in the US, such as Graham-Leach-Bliley Act (GLBA)~\cite{GLBA} for financial data or Health Insurance Portability and Accountability Act (HIPAA)~\cite{HIPA} for medical date provide consumers with limited protection scope when it comes to using AI and ML. 
Recently, the US introduced H.R. 8152-The American Data Privacy and Protection Act~\cite{HR8152}-to catch up with various amendments proposed in EU on AI/ML workloads by extending the covered algorithm to AI/ML space. Some states also patched their existing policies such as California Privacy Rights Act (CPRA)~\cite{CPRA} amendments to the CCPA~\cite{CCPA} or the Texas Data Privacy and Security Act (TDPSA)~\cite{texas}. 

GenAI’s remarkable capability to analyze data and perform complex analysis further amplifies privacy concerns.
There are clear privacy gaps in existing privacy policies as they relate to AI systems. Most of these policies and regulations were proposed pre-AI and even newer ones pre-Gen AI where there has been less information on privacy implications of these systems. Such gaps may pose significant challenges to privacy laws and they will need further amendments to address newer emerging privacy concerns.  

The recent AI Foundation Model Transparency Act~\cite {AIFMact} mandates FTC and NIST to establish standards for data sharing by foundation model deployers. This legislation aims to empower consumers to make welliinformed decisions when they interact with AI. The Algorithmic Accountability Act of 2023~\cite{AIAccAct} also requires companies to assess the impacts of the AI systems on consumer privacy and give users choices when they interact with AI systems.

\textbf{Future policy.} The previous regulations mostly require keeping the data in its location and restrict its processing based on intended collection purposes. GL is a potential solution to facilitate compliance with these privacy laws. 
In the near future, we anticipate a growing alignment between technological advancements -- especially in artificial intelligence (AI) and FL -- and evolving regulatory frameworks. This alignment will likely follow several key pathways.

First, there is a need for 
regulation specialized for AI use cases and application domains. 
Rather than relying on adapting existing frameworks like GDPR or HIPAA, regulatory bodies are expected to introduce laws specifically tailored to AI. These regulations will address the entire AI model lifecycle, covering aspects such as data collection, model training, decision-making, and retraining. The goal will be to ensure greater accountability, fairness, and transparency in AI systems. Additionally, the rise of generative AI requires a stronger focus on regulating not just the input data but also the output of AI systems. Transparency around how these generative systems, including large language models, are trained, the datasets they rely on, and how their outputs might impact privacy or perpetuate or amplify biases will become increasingly important. Moreover, governments can mandate real-time monitoring and reporting tools to ensure that AI systems maintain compliance with privacy regulations as they evolve dynamically.

Second, FL is poised to emerge as a valuable tool for compliance. As its use becomes more widespread, regulators may view it as a method that aligns with privacy-by-design principles. FL could be explicitly incorporated into compliance standards, particularly those influenced by GDPR-like data minimization principles, i.e., limit the collection of personal information to what is essential for a defined objective~\cite{GDPR}. By leveraging a decentralized approach, companies may be able to innovate while ensuring compliance with data privacy laws.


Overall, FL aligns well with privacy regulations like GDPR and HIPAA. It adheres to GDPR's principles of data minimization and purpose limitation. By keeping sensitive data on local servers, it significantly reduces the amount of sensitive information being processed centrally. Only model updates, which are typically less sensitive than raw data, are shared with the central server. FL systems can also be designed to collect only the data necessary for the specific purpose of model training.  Furthermore, 
individuals' rights over their data is supported by allowing users to opt-in or opt-out of the training process at any time. This aligns well with GDPR's emphasis on user control and consent. The decentralized nature of FL enhances data security, reducing the risk of large-scale data breaches. It embodies the principle of privacy by design which is a key requirement under GDPR.

Despite the above advantages, FL still faces some regulatory challenges. Data privacy laws differ across countries and regions, which can complicate global implementation of FL systems. Further, ensuring and demonstrating compliance across a distributed system can be complex and requires robust observability and documentation practices. The process of combining multiple models during aggregation can pose security risks, such as susceptibility to model poisoning attacks. In the next two sections, we discuss new and emerging privacy threats along with other defenses such as DP-SGD or secure aggregation which may be required to continue to protect user privacy in FL.

\section{Privacy Attacks}
\label{sec:attacks}

In this section, we first discuss the leading categories of privacy attacks against FL. We summarize this, with the most significant attributes of each attack type, in Table~\ref{tab:attack_summary}.

\subsection{Data Reconstruction Attacks}

\begin{figure}[htpb]
  \vspace*{-6mm}
  \centering
  \subfloat[BS 8 Ground truth]{
    \includegraphics[width=0.24\columnwidth]{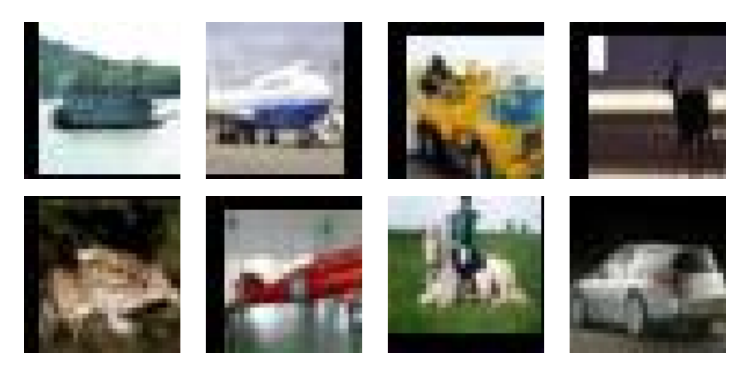}
    \label{fig:cifar8_gt}} \hspace*{10mm}
  \subfloat[BS 8 Reconstruction]{
    \includegraphics[width=0.24\columnwidth]{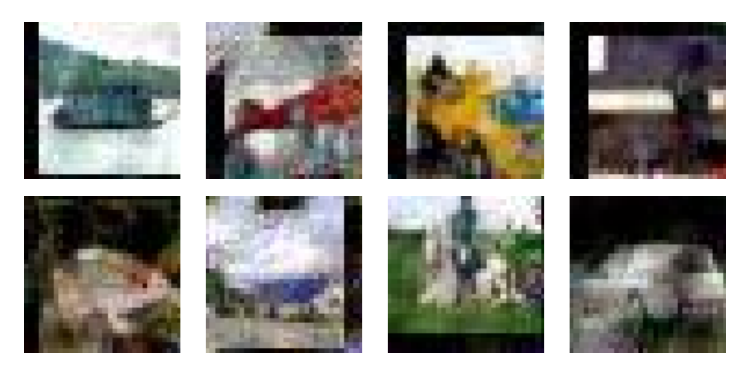}
    \label{fig:cifar8_rc}} \\
  \subfloat[BS 16 Ground truth]{
    \includegraphics[width=0.24\columnwidth]{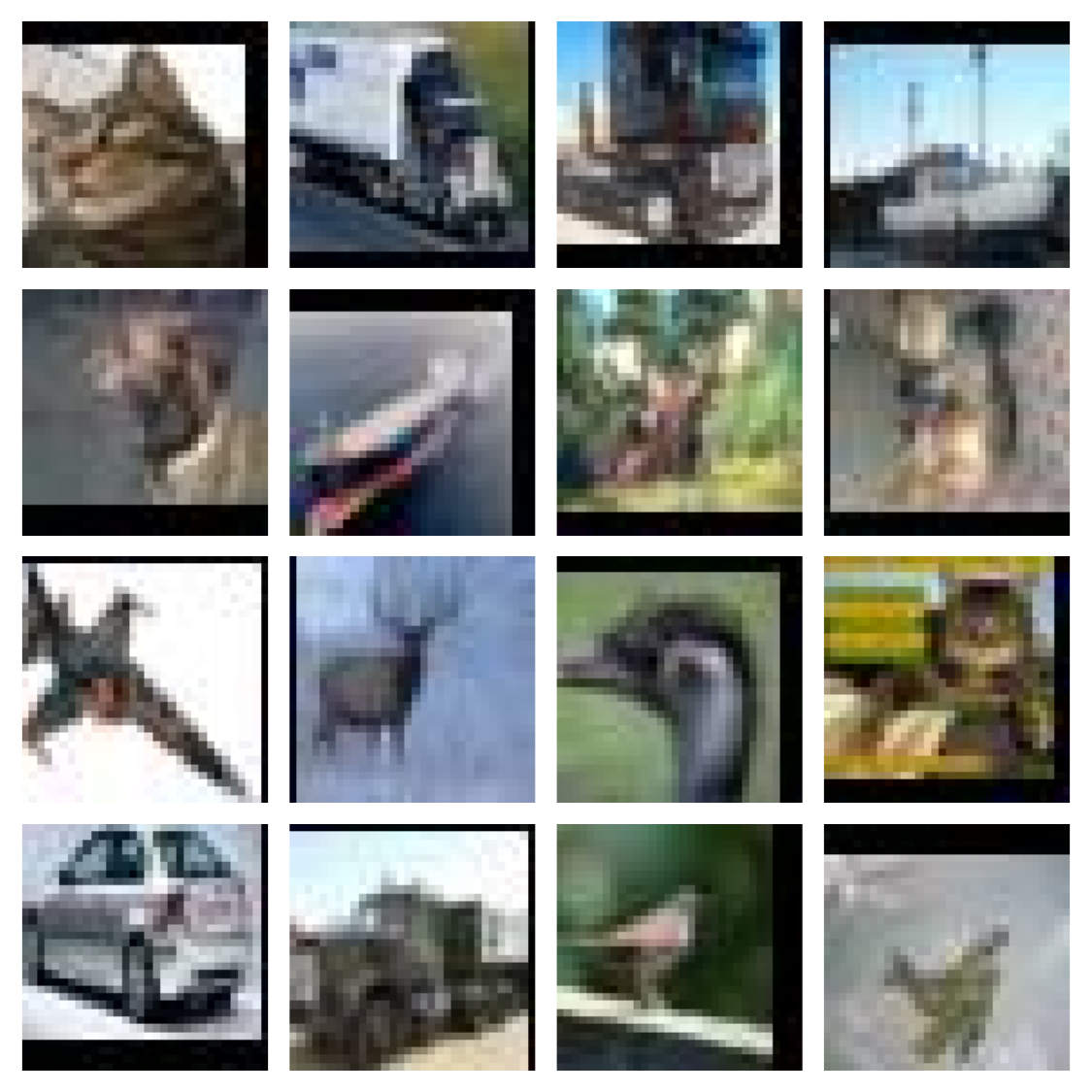}
    \label{fig:cifar16_gt}} \hspace*{10mm}
  \subfloat[BS 16 Reconstruction]{
    \includegraphics[width=0.24\columnwidth]{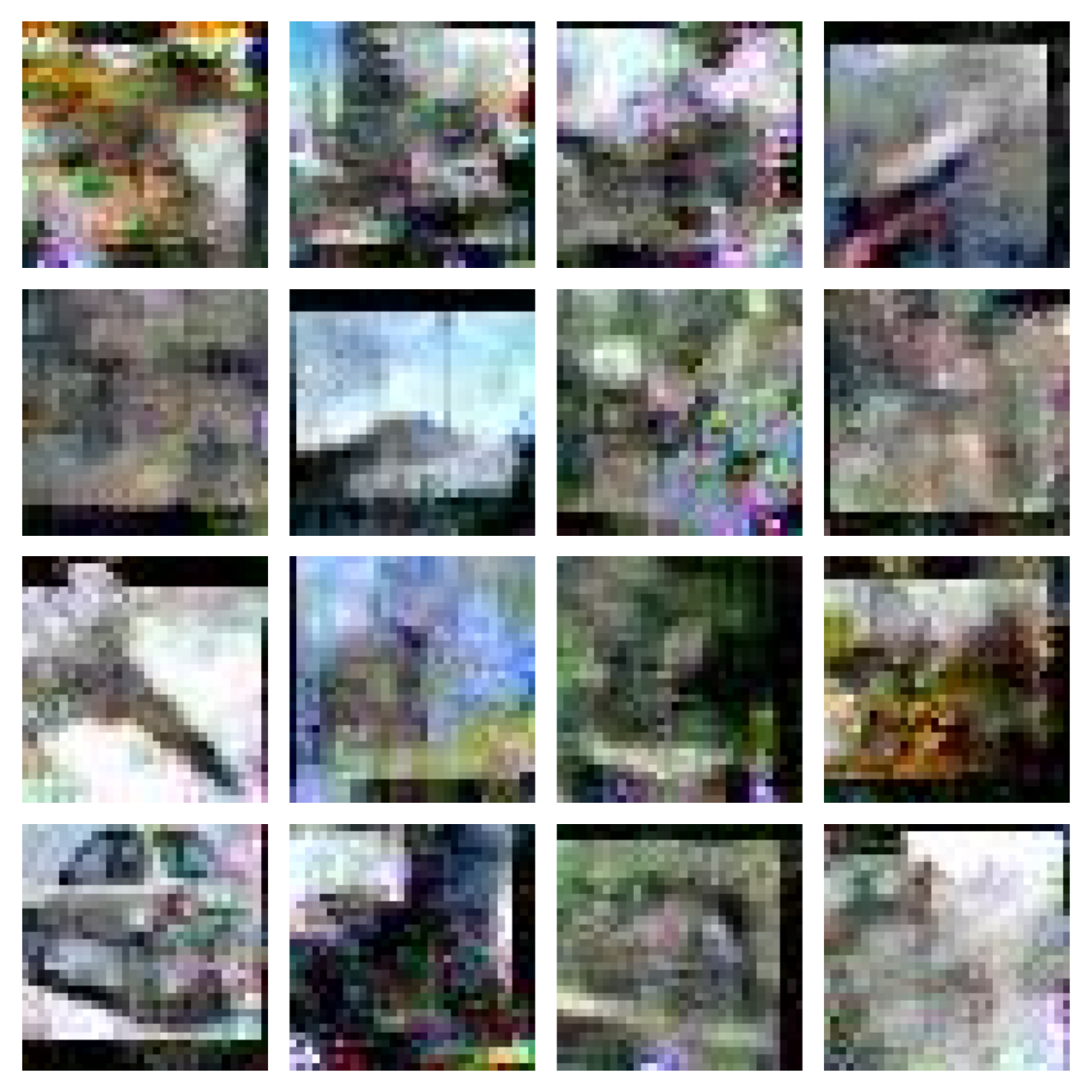}
    \label{fig:cifar16_rc}}
  \vspace*{-1mm}
  \caption{\label{fig:ig_cifar100} Images reconstructed by Inverting Gradients~\cite{geiping2020inverting}, an optimization attack, on a batch of 8 and 16 CIFAR-10 images in FedSGD. Ground truth images (a, c) are shown on the left, and reconstructions (b, d) are shown on the right. Images are visually recognizable for batch size 8, but become less recognizable at batch size 16.}
  \vspace*{-3mm}
\end{figure}

\begin{figure}[htpb]
  \vspace*{-1mm}
  \centering
  \subfloat[Ground truth]{
    \includegraphics[width=0.30\columnwidth,trim={20mm 20mm 20mm 20mm}]{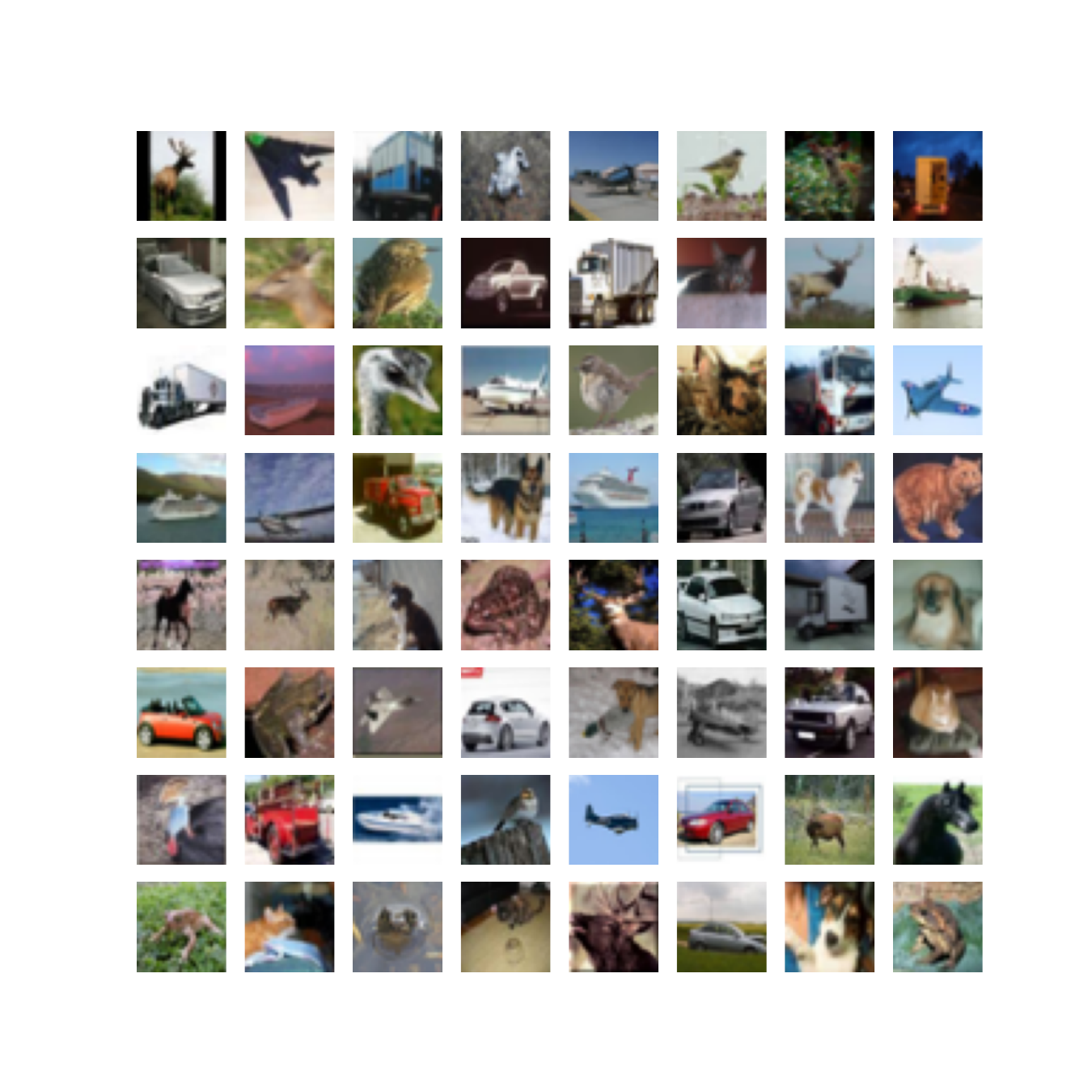}
    \label{fig:loki_gt_cifar}}\hspace*{10mm}
  \subfloat[Reconstruction]{
    \includegraphics[width=0.30\columnwidth,trim={20mm 20mm 20mm 20mm}]{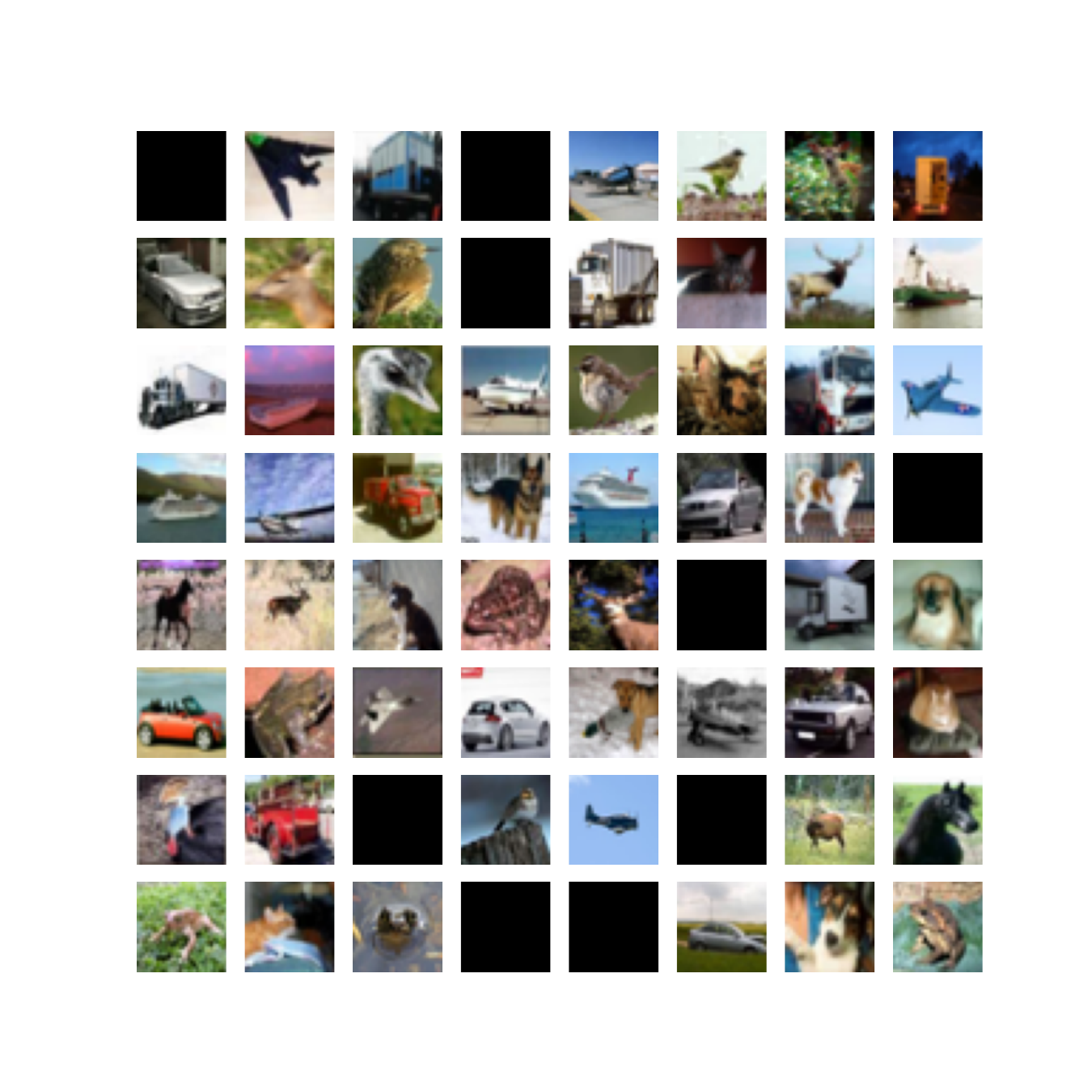}
    \label{fig:loki_recon_cifar}}
  \vspace*{-1mm}
  \caption{\label{fig:loki_examples_cifar} Images reconstructed using LOKI~\cite{zhao2023loki}, a linear layer leakage attack, on a batch of 64 CIFAR-10 images. FedAvg with 8 local iterations of mini-batch size 8 used. Out of 64 images, 55 are reconstructed successfully with high quality.}
  \vspace*{-4mm}
\end{figure}

\begin{table}[ht]
\scriptsize  
\centering
\caption{\label{tab:attack_summary} Summary of privacy attacks in federated learning}
\begin{adjustbox}{max width=\textwidth}
\begin{tabular}{p{2.3cm} p{4.2cm} p{2.3cm} p{2.4cm} p{3.6cm}}
\toprule
\textbf{Attack class} & \textbf{Summary} & \textbf{Sub-category} & \textbf{Citation} & \textbf{Requirements} \\
\midrule

\textbf{Data reconstruction} 
& \multirow{4}{4.2cm}{%
Attacker aims to directly reconstruct the client private data or a representation of the private data. 
Typically, this involves analyzing gradient updates or partial model parameters that inadvertently reveal features of the original input. 
With sufficient information, the attacker can refine these reconstructions to near-original fidelity, posing severe privacy risks—particularly in sensitive domains like healthcare. 
Factors such as model complexity, dataset size, and update granularity can further exacerbate this threat.}
& Gradient inversion
& \cite{zhu2019deep,zhao2020idlg,geiping2020inverting,yin2021see}
& \begin{tabular}[c]{@{}l@{}}Access to model and\\ individual updates.\end{tabular} \\

& 
& \begin{tabular}[c]{@{}l@{}}Linear layer\\ leakage\end{tabular}
& \cite{zhao2023loki,fowl2022robbing,boenisch2021curious}
& \begin{tabular}[c]{@{}l@{}}Modification to the\\ network parameters\\ and/or architecture.\end{tabular} \\

& 
& GAN-based
& \cite{hitaj2017deep,wang2019beyond}
& \begin{tabular}[c]{@{}l@{}}Clients target a\\ specific class by\\ sending malicious\\ updates.\end{tabular} \\

& 
& Other
& \cite{wen2022fishing,pasquini2021eluding,lam2021gradient}
& \begin{tabular}[c]{@{}l@{}}Varies based on attack.\\ Attacker may modify\\ the model parameters,\\ know training info, or\\ send different updates.\end{tabular} \\
\midrule

\textbf{Membership inference}
& \begin{tabular}[c]{@{}l@{}}Attacker infers if a sample was \\ used in the training of the model.\end{tabular}
& 
& \cite{song2019membership,nasr2019comprehensive,zari2021efficient,melis2019exploiting}
& \begin{tabular}[c]{@{}l@{}}Most require access to training data \\ from the overall distribution. Some \\ attacks require white-box access to \\ the model or the ability to send \\ malicious updates.\end{tabular} \\
\midrule

\textbf{Property inference}
& \begin{tabular}[c]{@{}l@{}}Attacker learns about sensitive \\ properties within the training set \\ (e.g., race, gender, age).\end{tabular}
& 
& \cite{fredrikson2015model,mehnaz2022your,dibbo2023model,ganju2018property}
& \begin{tabular}[c]{@{}l@{}}Attacker has black-box access \\ to the model and can access \\ the output. They may also \\ possess other attributes beyond \\ the targeted one.\end{tabular} \\
\midrule

\textbf{Model extraction}
& \begin{tabular}[c]{@{}l@{}}Attacker wants to steal \\ functionality of a model. This can \\ be the parameters, hyperparameters, etc.\end{tabular}
& 
& \cite{tramer2016stealing,wang2018stealing,orekondy2019knockoff}
& \begin{tabular}[c]{@{}l@{}}Attacker needs black-box access \\ to the model in order to query \\ it and observe the outputs.\end{tabular} \\
\bottomrule
\end{tabular}
\vspace*{-5mm}
\end{adjustbox}\end{table}

\subsubsection{Optimization-based.}
Optimization (a.k.a. gradient inversion) approaches have shown great success in leaking data from individual updates, especially with smaller batch sizes. These attacks typically operate under the threat model of an honest-but-curious server or an external attacker that has access to the model and individual gradients from each client. With only this information, the attacker initializes some dummy data (i.e., with gaussian noise) and computes the gradient of the data.
\begin{equation}\label{eq:optim}
    x^* = \arg \min_{x}||\nabla L(x, y, \theta)-\nabla W||_2
\end{equation}
\noindent
An optimizer minimizes the difference between the generated gradient $\nabla L(x, y, \theta)$ and the ground truth gradient $\nabla W$ (benign client update). In the above equation, $x$ is the dummy data, $x^*$ is the reconstructed data, $L$ is the loss function, $y$ is the label (which can be known beforehand or solved for~\cite{zhao2020idlg}), and $\theta$ is the model parameters.

More recent optimization approaches~\cite{geiping2020inverting,yin2021see} work under the assumption that user labels are known prior to optimization. Typically, these labels are directly retrieved through a zero-shot solution without using optimization approaches~\cite{zhao2020idlg,yin2021see}. Furthermore, regularizers and strong image priors specific to image data are often used to guide optimization results~\cite{geiping2020inverting,yin2021see,hatamizadeh2023gradient,usynin2023beyond}. These can also result in image artifacts typical of an image class, but not in the actual training image. These approaches have shown surprising success with image data on smaller batch sizes. However, as batch sizes increase, the fraction of images recovered decreases along with the reconstruction quality and the number of iterations required for the optimization also increases. One reason stated by~\cite{zhu2019deep} was that regardless of the order of images in the batch, the gradient will remain the same. Having multiple possible permutations then makes the optimization more difficult. Another fundamental reason is that a larger batch size means more images and more variables for optimization. Figure~\ref{fig:ig_cifar100} shows Inverting Gradients~\cite{geiping2020inverting} reconstructions of batch size 8 and 16 on CIFAR-100, demonstrating this property of lowering quality with larger batch sizes.

Both secure aggregation and FedAvg also pose a particularly difficult challenge for optimization attacks. For secure aggregation, the aggregated updated can be thought of as large-batch consisting of the batch images from all clients. This directly exacerbates the difficulty of reconstructing larger batch sizes. FedAvg, on the other hand, adds another layer of unknowns for optimization since the intermediate model updates are unknown to the attacker. While~\cite{geiping2020inverting} discuss attacking FedAvg, the attack is only shown with a very small local dataset size of up to 8 images on CIFAR-10. Furthermore, it needs to be investigated how different model architectures, such as transformers~\cite{hatamizadeh2022gradvit}, impact the success of gradient inversion attacks.

\subsubsection{Linear layer leakage.}
\begin{figure}[!t]
\begin{center}
\includegraphics[width=0.6\columnwidth]{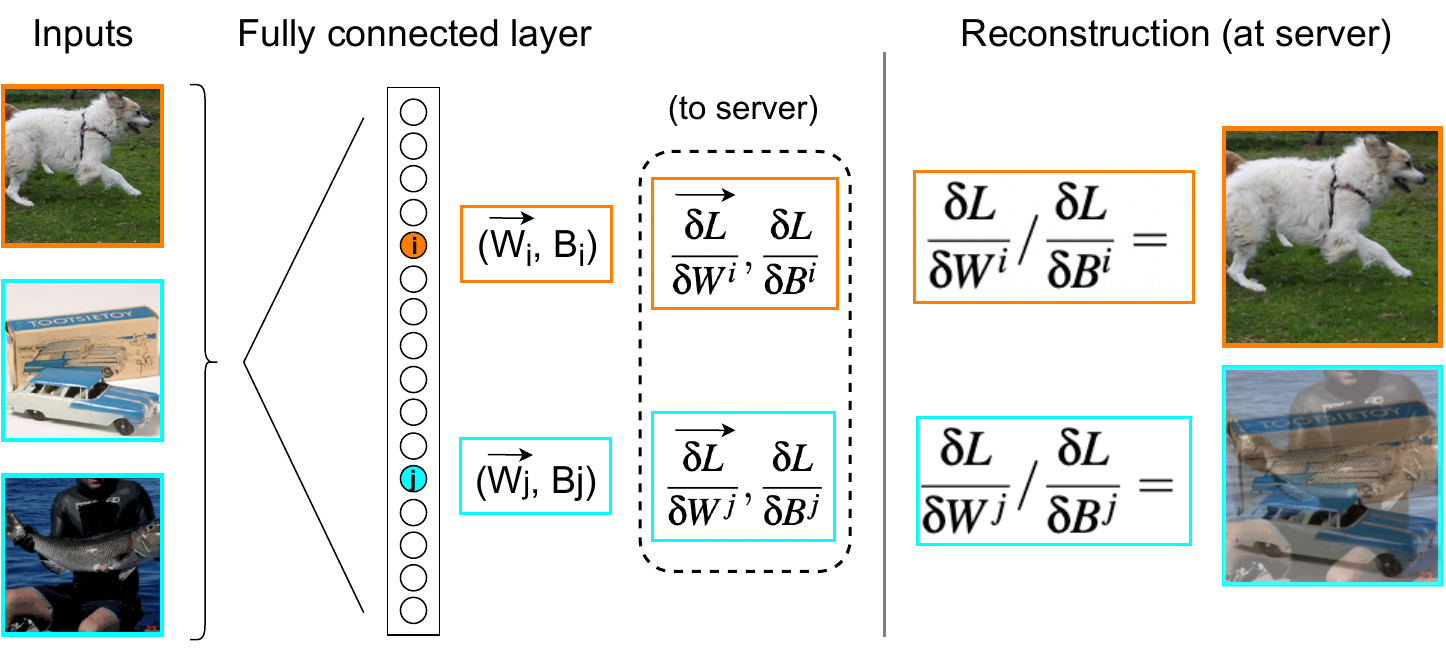}
\end{center}
\vspace*{-3mm}
\caption{\label{fig:linear-leak-method} Using the weight gradient $\frac{\delta L}{\delta W}$ and bias gradient $\frac{\delta L}{\delta B}$ of a fully connected layer to reconstruct the inputs. Neuron $i$ is only activated by a single image, while $j$ is activated by two. As a result, the reconstruction of neuron $i$ is correct while $j$ is a combination of images.}
\vspace*{-3mm}
\end{figure}

Linear layer leakage attacks are a sub-class of analytic attacks that modify FC layers to leak inputs. Using the weight and bias gradients of an FC layer to leak inputs was discussed in~\cite{phong2017privacy,fan2020rethinking}. When only a single image activates a neuron in a fully connected layer, the input to that layer can be directly reconstructed with:
\begin{equation}\label{eq:1}
    x^i = \frac{\delta L}{\delta W^i} / \frac{\delta L}{\delta B^i}
\end{equation}
\noindent
where $i$ is the activated neuron, $x^i$ is the input that activates neuron $i$, and $\frac{\delta L}{\delta W^i}$, $\frac{\delta L}{\delta B^i}$ are the weight gradient and bias gradient of the neuron respectively. This idea forms the basis for several reconstruction attacks~\cite{boenisch2021curious,fowl2022robbing}. Figure~\ref{fig:linear-leak-method} shows the basic process of leaking images through an FC layer. 

When the fully-connected layer is placed at the start of a network, the data reconstructed from that layer would be the input data. This reconstruction is exact, as opposed to the optimization approaches which function as estimations. However, inputs are only reconstructed exactly when a single data sample activates that neuron. If more than one input activates the neuron, the weight and bias gradients of these inputs will contribute to the batch gradient. When the gradient division of Equation~\ref{eq:1} is done to retrieve the input, the resulting reconstruction would be a combination of all contributing images, a case of failed attack. 

To alleviate this problem,~\cite{fowl2022robbing,boenisch2021curious,zhao2023loki} use malicious modification of the parameters in the FC layer. For~\cite{boenisch2021curious}, trap weights were introduced, initializing the weights randomly to be half positive, half negative. In order to ensure that neuron activation is less common, the negative weights come from a larger negative magnitude range than the positive weights. They also discuss the use of convolutional layers to push the input image forward, allowing the attack to function on models starting with convolutional layers followed by fully-connected layers. However, one of the main problems of the method lies with scalability. Even if the size of the FC layer increases proportionately with an increasing total number of images, the leakage rate decreases. On the other hand, Robbing the Fed (RtF)~\cite{fowl2022robbing} introduced another approach with higher leakage rate called ``binning", where the weights of the FC layer would measure some known continuous CDF of the input data such as image brightness. The bias for each neuron then serves as a different cutoff, allowing only inputs with a high enough value to activate it. The goal of this method would be that only one input activates each ``bin", where the bin is defined as the activated neuron with the largest cutoff (for ReLU, the largest negative bias)\footnote{The bin biases are set as negative. The weights are positive and so the negative bins are used to prevent ReLU activation.}. For any case where only one input activates a bin, it can then be reconstructed as
\begin{equation}\label{eq:2}
    x^i = (\frac{\delta L}{\delta W^i} - \frac{\delta L}{\delta W^{i+1}}) / (\frac{\delta L}{\delta B^i} - \frac{\delta L}{\delta B^{i+1}})
\end{equation}
\noindent
where $i$ is the activated bin and $i+1$ is the bin with the next higher cutoff bias. For Equation~\ref{eq:2} to hold true, the attack requires the use of two consecutive FC layers. The first layer is used to leak the inputs using Equation~\ref{eq:2} and the second FC layer maintains the requirement that $\frac{\delta L}{\delta B^i}$ and $\frac{\delta L}{\delta W^i}$ are the same for any neuron that the same input activates. This is achieved by having the same weight parameters connecting each neuron of the first FC layer to the second FC layer. For example, if the first FC layer has $1024$ units and the second has $256$, the weights connecting them would have 
a dimension of $1024\times256$. The above property indicates that every row of the weight matrix is equivalent, e.g. $[0, :]=[1, :] = \dots = [1023, :]$. 

\noindent \textbf{FedAvg}. While the previous method works for FedSGD, for FedAvg, the model changes during local iterations and this prevents the reconstruction attack. As a result,~\citet{fowl2022robbing} proposed the sparse variant of the attack which uses an activation function with a double-sided threshold (e.g., Hardtanh) such as:
\begin{equation}\label{eq:act_func}
    f(x) = \protect\begin{cases} 0 & x \leq 0 \\ 
    x & 0 \leq x \leq 1 \\ 
    1 & 1 \leq x  \protect\end{cases}
\end{equation}
\noindent With this activation function, only when the input is between $0$ and $1$ will there be a non-zero gradient. 

Using this activation function, neuron activation will be sparse (i.e., images will only activate a single neuron). However, this range between 0 and 1 for the non-zero gradient is fixed for all neurons. Since RtF's approach sets up neuron biases following a distribution of the images, the weights and biases of the FC layer will need to be adjusted to follow the new non-zero gradient range. This requires scaling the magnitude of these parameters based on the distance between the subsequent neuron biases. Consider that the weights originally measure the average pixel brightness. In this case, all weights would originally be set to $\frac{1}{N}$, where $N$ is the total number of pixels. Then, the weights and biases are rescaled as:
\begin{equation}\label{eq:param_scale}
\begin{split}
W^*_i = \frac{W_i}{b_{i+1} - b_i} \text{  } , \text{  } b^*_i = \frac{b_i}{b_{i+1} - b_i}
\end{split}
\end{equation}
\noindent where $W^*_i$ and $b^*_i$ are the scaled weights and biases of neuron $i$, respectively, and $b_{i+1} - b_i$ is the distance between adjacent biases in the original distribution. This process uses the same distribution as the FedSGD case to setup the initial biases, while incorporating the fixed range of the new activation function by scaling the parameters. In FedAvg, after the clients send the updated model parameters, the server computes a ``gradient" as:
\begin{equation}\label{eq:FedAvg_grad}
    \nabla W_{FedAvg} = \Theta_{t+1} - \Theta_{t} 
\end{equation}
where $\Theta$ is the model parameters for the securely aggregated model and $\nabla W_{FedAvg}$ the computed gradient.

This FedAvg attack~\cite{fowl2022robbing} works well in the context of single client attacks. However, attacking secure aggregated gradients becomes a problem as the total number of images in the gradient increases multiplicatively to the number of clients. In a recent attack, ~\citet{zhao2023loki} (LOKI) finds that numerical precision poses a problem for the FedAvg linear layer leakage attack when a large number of clients are attacked. Through the use of convolutional layers and a convolutional scaling factor (CSF), LOKI is able to scale to an arbitrary number of clients while maintaining a high leakage rate. Furthermore, the CSF improves on the multiple-image reconstruction problem, only allowing images within the same local mini-batch to activate the same units. Notably, this allows the leakage rate of FedAvg attacks to {\em surpass} the leakage rate in FedSGD (where FedAvg is typically thought to be harder to attack than FedSGD). Figure~\ref{fig:loki_examples_cifar} shows the reconstructions from LOKI on a total of 64 images from CIFAR-100 in FedAVG (split over 8 local iterations). LOKI achieves a high leakage rate, with $\frac{55}{64}$ images reconstructed with high quality.

Linear layer leakage reconstruction quality is typically higher quality than optimization-based attacks and is not impacted by the number of clients. However, it also requires malicious modification of the model parameters and/or architecture, resulting in more detectable attacks. Furthermore, the attack also adds a resource overhead to the total model size. This problem is exacerbated by secure aggregation, as the size of the attack layer needs to scale to keep the attack effective. This problem along with the use of sparse tensors to alleviate it are discussed by~\citet{zhao2023resource}.

\subsubsection{GAN-based.} Generative networks were introduced to attack collaborative learning in~\cite{hitaj2017deep,wang2019beyond}. However, contrary to optimization-based attacks and linear layer leakage, the attack does not aim to directly reconstruct exact training images but instead reconstructs representations of a targeted class which the attacker does not hold any images of. In the attack, an adversary (a client in collaborative or federated learning training) trains a generative network in conjunction with the benign collaborative model. For a single training round, the attack process goes as follows:
\begin{enumerate}
    \item The adversary downloads the model (or a portion of parameters in certain forms of collaborative training).
    \item Images for the targeted class are generated by the generator and the generator is updated based on the discriminator (the global model).
    \item Images for the targeted class are generated by the updated generator and mislabeled.
    \item The adversary trains on the set of mislabeled images and any actual training images.
    \item The adversary uploads the update to the parameter server.
\end{enumerate}
Both training the generator and training on mislabeled images are critically important to the attack process. Firstly, training the generator is crucial since the ultimate goal is to be able to generate representations of a targeted class. Secondly, the process of including mislabeled generated images in order to influence the training process is also very important. Without this additional influence on training, the authors note that the generated image quality does not converge to a visible reconstruction.

One main strength of the mention is also in the ability to work through differential privacy. Specifically, the attack relies on the fact that the client model is able to learn and improve the accuracy. As long as the model is learning, the generator is also able to learn to reconstruct better images. This makes the GAN attack remarkably resilient to (at least record-level) differential privacy. As long as the privacy budget $\epsilon$ allows the model to continue learning, the attack works. When too restrictive of an $\epsilon$ is used, the attack fails but the model is also unable to learn.

Another strength of the GAN attack is that it can be employed through any malicious client as opposed to a parameter server. Optimization-based attacks require the observation of individual gradient updates along with the initial model. While clients receive the initial model state, the observation of individual gradient updates from other clients would not happen in the vanilla protocol. Linear layer leakage on the other hand can only be enacted by the server.  Modification of the model parameters is required, and this can only be done by the parameter server. Compared to both these methods, only the GAN attack can be easily employed by a client. This larger attack surface also points to potentially greater privacy risks in collaborative training compared to even centralized training. It should also be noted that the attack only generates representations instead of actual reconstructions. However, this can still pose a great privacy risk depending the content of the private client training data.

\subsubsection{Other attacks.}
While we have previously discussed the major categories of optimization, linear layer leakage, and GAN-based attacks for data reconstruction, other attacks have also been proposed that leak data differently or assist existing methods in attacking more challenging scenarios such as secure aggregation. 

In order to tackle secure aggregation, a gradient suppression attack~\cite{pasquini2021eluding} was proposed whereby a malicious server sends customized model parameters to different clients such that only a single client would return a non-zero gradient update. This attack used a dead ReLU trick where the ReLU activation functions of a layer using were set to 0.
\begin{equation}\label{eq:relu_act}
    ReLU(x) = \protect\begin{cases} 0 & x \leq 0 \\ 
    x & x > 0\end{cases}
\end{equation}
For a ReLU activation, any input below 0 returns a value of 0. Given the constant function when $x\leq0$, this ultimately results in the gradients $\frac{\delta L}{\delta W}=0$ and $\frac{\delta L}{\delta b}=0$. Therefore, ensuring the input to the ReLU function is negative (which can be as simple as having weights $w$ as 0 and the bias $b$ as negative for a fully-connected layer) is a sufficient condition for achieving zeroed gradients. Using this method, a malicious server will then send a malicious model with dead-ReLUs to all clients except for a single targeted client. After secure aggregation, the aggregated update would only comprise of a single non-zero update from the targeted client allow for a single-update attack. Another work~\cite{lam2021gradient} aims to disaggregate the updates with the same goal of breaking aggregation. This method utilizes the additional training summary information, in particular the client participation rate, to allow the server to reconstruct the user participation matrix through observation of the aggregated updates over multiple training rounds. The attack works particularly well with a malicious server when the a fixed model is sent during each round and the clients send the same updates. When the server does not modify the training process (honest-but-curious), user participation reconstruction becomes more difficult. With a smaller batch size and a larger number of local epochs or local dataset size, reconstruction begins to fail.

In~\citet{wen2022fishing}, attackers attack a target image in an arbitrarily large batch by manipulating the model parameters to magnify the update of only the single image. Magnification is done in two steps: the magnification of the target class followed by a specific feature for a target image. In order to estimate a boundary for the features, the server uses several rounds of additional FL training for setup. This requires for the batch data to be the same across training rounds. \citet{kariyappa2022cocktail} formulates the data reconstruction attack as a blind source separation problem. Given the observation that inputs to a linear layer can be directly reconstructed through the gradients, solving for the inputs is similar to solving for a unknown variables given a set of linear equations.

These attacks are all capable of breaking some levels of aggregation. \cite{pasquini2021eluding,lam2021gradient} directly break aggregation and other data reconstruction attacks to work on top of them. ~\cite{wen2022fishing,kariyappa2022cocktail} both function on individual and aggregate updates. Each of these methods have their own strengths and weaknesses. \citet{pasquini2021eluding} requires sending different models to different clients but allows an attacker to single out a single update. \citet{lam2021gradient} requires multiple training rounds and additional knowledge in addition to the client update. However, it can essentially disaggregate aggregate updates over time. \citet{kariyappa2022cocktail} can attack batch sizes up to 1024, but also requires malicious modification of the network parameters. \citet{wen2022fishing} targets a single image in an arbitary update size, but requires malicious modification of network parameters along with a few rounds of setup.

\subsection{Membership Inference Attacks}
\label{sec:attacks_mem_inf}

\begin{figure}[]
    \centering
    \includegraphics[width=0.6\textwidth]{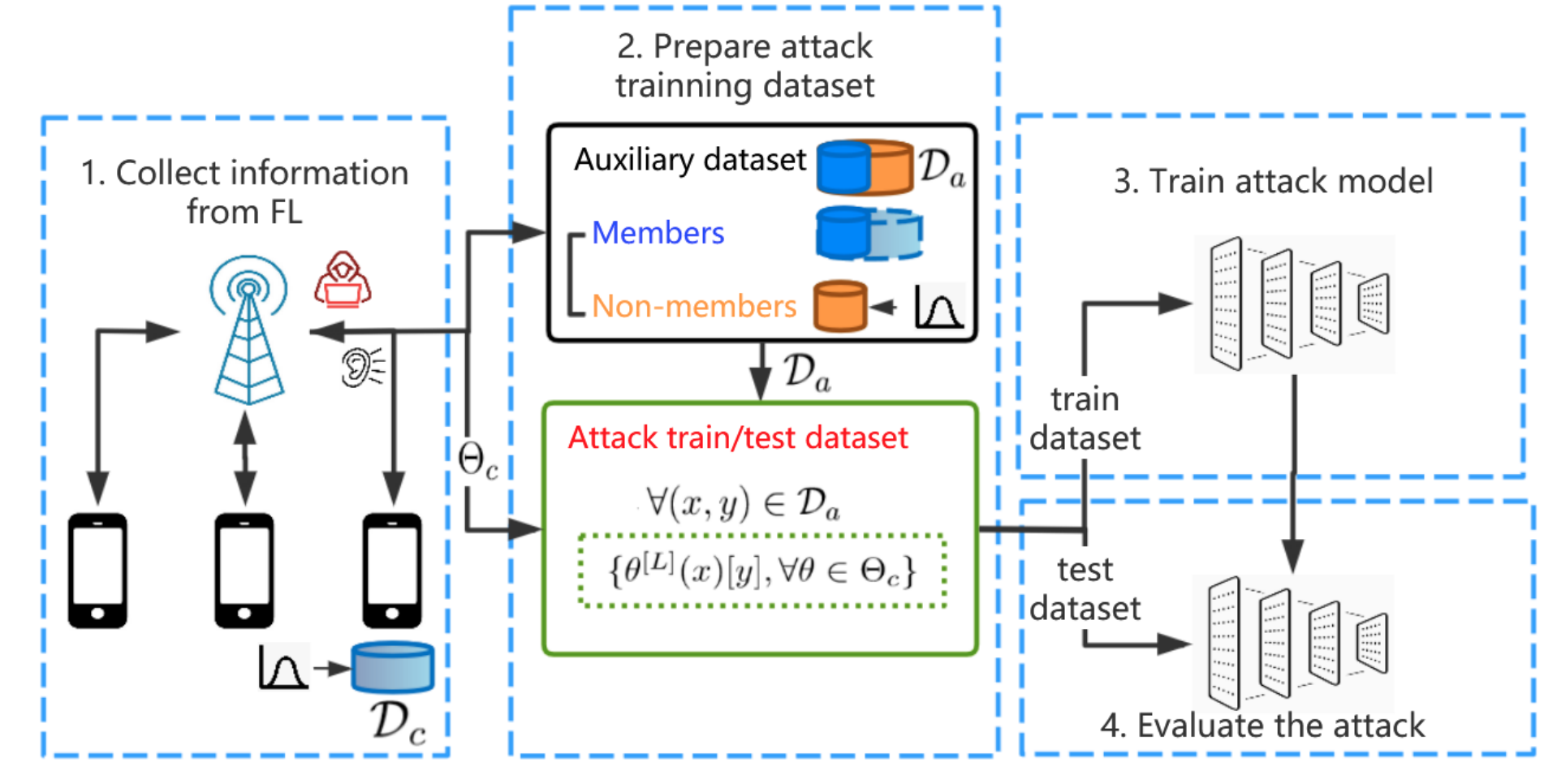}
    \caption{Membership inference attack workflow from~\cite{zari2021efficient}. The attacker first collects information (e.g. parameter updates) from clients, then uses auxiliary dataset to train shadow models which can mimic the whole federated learning procedure, so that one attack model can be trained using the information collected from shadow models. Lastly, the attacker feeds the information collected from real clients into the attack model to perform membership inference attacks.}
    \label{fig:membership_inf}
    \vspace{-5mm}
\end{figure}

The goal of membership inference attacks is to infer whether any particular data instance has been used in the training of a specific model. If a particular data instance has been used in the training, this instance is called a  \textbf{member}, otherwise it is a \textbf{non-member}. Knowing the membership of one particular data point could result in revealing private information, for example, if someone's data is known to be in a cancer dataset (used to train a cancer prediction model), then it is highly likely that this particular person has cancer. Figure~\ref{fig:membership_inf} summarizes the process of membership inference attacks.

\subsubsection{With access to some training data.}

\citet{nasr2019comprehensive} presented the very first analysis of the membership inference attack against neural networks under FL with access to some training members and white-box access to trained neural networks. The adversary can be either the central server or one of the participants in the FL framework. They also presented the definition of passive attacker and active attacker, where the passive attacker does the training normally and the active attacker breaks the FL protocols to improve the effectiveness of MI attacks.

For passive attackers, they showed that attackers can take advantage of the gradients, activation maps, prediction vectors, loss, and true label of a single instance using local model of each user at different epochs during the training process and perform MI attacks. An attack model is trained using some known members. 

On the other hand, two active attacker strategies were proposed. The first is the \textbf{gradient ascent attacker}, which means that the attacker will apply gradient ascent on targeted samples. By taking the action of gradient ascent, the loss of the targeted samples will increase. For member instances, their loss will be abruptly decreased by one client, thus allowing distinguishability from non-members. Both a central server adversary and a client adversary can utilize this strategy. The second active attack method is called the \textbf{isolating attacker}. The attacker can isolate one special target participant by creating a separate copy of the aggregated model. This is equivalent to training a model on a single machine using the data from the targeted client. To use this isolation attack, the adversary would be a malicious central server. 

\citet{zari2021efficient} proposed a computationally more efficient passive membership inference attack under FL which also requires some known members. Their attack uses as the feature vector the probabilities of the correct label under local models at different epochs, namely $\big\langle F_{\theta_{C_i}^{(1)}}(x)_{y}, F_{\theta_{C_i}^{(2)}}(x)_{y}, \cdots, F_{\theta_{C_i}^{(T)}}(x)_{y}\big\rangle$, where $T$ is the number of communication rounds. An attack model is trained to predict membership using known members and known non-members (which come from validation). Since the feature vector contains only one number from one epoch, this attack requires significantly less computational resources than the Nasr's attack which uses information such as gradient and activation maps. The authors showed that using this probability feature vector can achieve higher membership inference attack accuracy than Nasr's attack on CIFAR-100 dataset with AlexNet and DenseNet, but lower accuracy on Purchase dataset.

\citet{melis2019exploiting} identified membership leakage when using FL for training a word embedding function, which is a deterministic function that maps each word to a high-dimensional vector. Given a training batch (composed of sentences), the gradients are all $0$'s for the words that do not appear in this batch, and thus the set of words used in the training sentences can be inferred.  The attack assumes that the participants update the central server after each mini-batch, as opposed to updating after each training epoch.

\subsubsection{Without access to any training data.}

\begin{figure}[]
    \centering
    \includegraphics[width=0.8\textwidth]{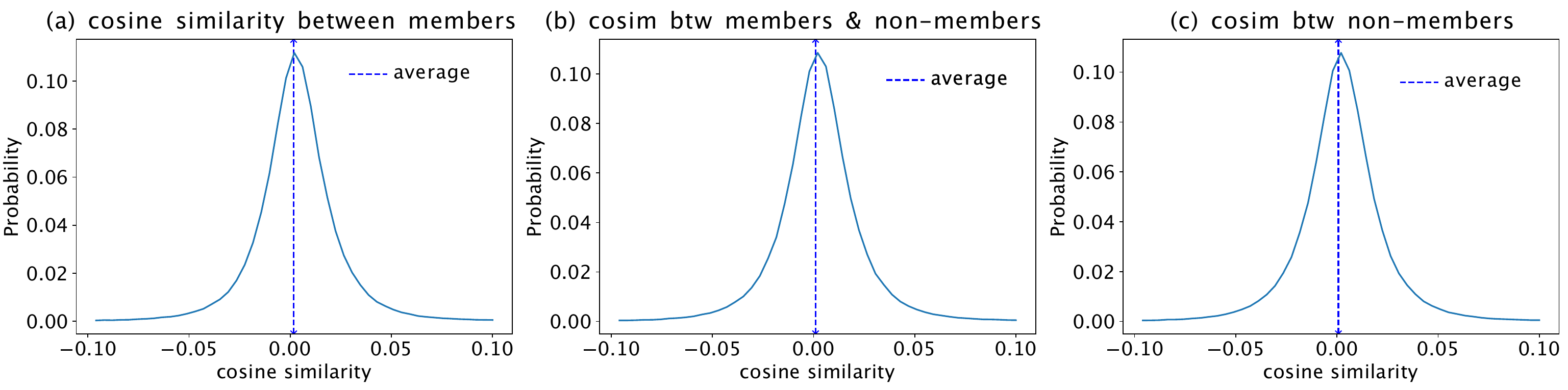}
    \caption{Representative example of orthogonality of gradients of distinct instances (at epoch 250, AlexNet, CIFAR-100 dataset) from ~\cite{li2022effective}: (a) Distribution of cosine similarity (cosim) between gradients of two distinct member instances. (b) Distribution of cosine similarity between gradients of one member instance and one non-member instance. (c) Distribution of cosine similarity between gradients of two distinct non-member instances. }
    \label{fig:instance_cos_dis}
    \vspace*{-4mm}
\end{figure}

\citet{li2022effective} proposed a novel membership inference attack specifically against FL {\em without} requiring access to private training data. The key insight is that large overparameterized neural network models usually generalize well and the gradients of large overparameterized neural network models statistically behave like high-dimensional independent isotropic random vectors (shown in Figure \ref{fig:instance_cos_dis}). Thus, they reformulated the task of membership inference attacks under FL as the following question: given a set of (largely unknown) orthogonal vectors, which is the parameter update coming from a particular client, does this set contain the gradients of a specific target instance?

One straightforward solution to this question is to test the cosine similarity between the gradients of the target instance and the parameter update from one particular client. If the cosine similarity is significantly higher than $0$, then the attacker can claim with high confidence that this target instance is used in the training of this particular client, hence successful membership inference. 

From another perspective, if $S = a+b+c$ is a sum of orthogonal vectors, then $\Vert S \Vert^2_2 - \Vert S - a \Vert^2_2 = \Vert a \Vert^2_2$, otherwise for another vector $f$ orthogonal to $a,b,c$ we have $\Vert S \Vert^2_2 - \Vert S - f \Vert^2_2 = - \Vert f \Vert^2_2$.
These equations are easy to understand if we rotate the (orthogonal) vectors to form a canonical orthogonal basis $(a',0,0,\ldots), (0,b',0,\ldots),(0,0,c',\ldots)$, with $a',b',c' \in \mathbb{R}\backslash \{0\}$.  Each gradient vector in the sum $S$ is now associated with a unique canonical orthogonal basis. Since the $L_2$ norm is invariant under basis rotations (for the same reason the length of a vector is invariant under rotation), it now becomes clear why subtracting a vector $v$ of the orthogonal basis from $S$ reduces the $L_2$ norm of $S$ (if $v$ is in the sum) or increases the sum's $L_2$ (if $v$ is not in the sum). If the $L_2$ norm is reduced, the attacker could claim that the target instance is a member, otherwise it is a non-member.

\subsubsection{Membership inference attacks from centralized setting.}

It is worth noting that all the membership inference attacks that do not require \textit{shadow models} can be applied to FL setting as well. \textit{Shadow models} are a commonly used technique in membership inference attacks, and allow the attacker =to train models similar to the target model using data drawn from the same distribution of the training data of the target model. Thus, shadow models enable attackers to collect any information necessary for membership inference, for example, what the loss would be for one instance if this instance is not included in training.

\citet{yeom2018privacy} presented a theoretical analysis on privacy leakage based on prediction loss. They showed that the generalization gap between training accuracy and testing accuracy could serve as a lower bound for MI attack accuracy, as the attacker predicts the member if and only if the model's prediction is correct. Besides, they proposed a threshold attack based on prediction loss, and the attacker predicts member if and only if the prediction loss is lower than the threshold, since the loss of members is generally lower than the loss of non-members. 

Inspired by \citet{yeom2018privacy}, \citet{song2019membership} proposed to adjust the prediction loss by a class-dependent value, noticing that some classes are harder to correctly classify than other classes. In addition, \citet{song2019membership} proposed using a modified prediction entropy to predict membership. The modified prediction entropy is calculated using this formula: 
$ Mentr(F_{\theta}(x),y) = -(1-F_{\theta}(x)_y)\log(F_{\theta}(x)_y) - \sum_{i \neq y} F_{\theta}(x)_{i}\log(1-F_{\theta}(x)_{i})$.

\citet{jayaraman2021revisiting} suggested one MI attack based on Gaussian noise. The intuition is that for a member instance, the prediction loss should increase after adding random noise. This MI attack adds multiple different random noise to the given instance and count how many times the prediction loss of the noisy instance is higher than the prediction loss of the original instance. The given instance is predicted to be a member if the count is beyond a threshold set by the attacker. 

\citet{hui2021practical} exploits the model's prediction in a special ``set'' fashion. The adversary is aware of a set $M$ of members, and $N$ of non-members, and compute $d_1 = D(M\cup \{x\}, N)$ and $d_2 = D(M, N\cup\{x\})$, where $D$ denotes the \emph{Maximum Mean Discrepancy (MMD)}~\cite{fortet1953convergence} of two distributions of features extracted from the prediction vectors.  The intuition is that if $x$ is a member, then we tend to have $d_1 > d_2$, and if $x$ is non-member, we tend to have $d_1 < d_2$. This attack only provides predicted membership, which means that it is infeasible to evaluate its effectiveness at a low false positive rate.

\subsection{Property Inference Attacks}
\label{sec:attacks_model_inv}


\citet{fredrikson2015model} proposed the very first black-box property inference attack where the goal is to infer sensitive attributes of a given target record which is used in the training of a target model. This black box property inference attack assumes that the adversary can obtain the model’s predicted label, has knowledge of the attributes of a targeted record (including the true label) except the sensitive attribute, has access to the marginal priors of the attributes, and also to the confusion matrix of the target model (which is usually not known by the attacker). The adversary queries the target model by varying the sensitive attribute and obtains the predicted labels. Assuming that there is $k$ different values for the single sensitive attribute that the attacker wants to know, after querying the model multiple times with $k$ different possible values while keeping the other known attributes unchanged, the adversary computes $C(y,y')*p_i$ for each possible sensitive attribute value, where $C(y,y') = Pr [F(x) = y' | y]$ (which comes from the confusion matrix) and $p_i$, which is the marginal prior of i-th possible sensitive attribute value. Finally, the attack predicts the sensitive attribute value for which the $C(y,y')*p_i$ value is the maximum. 

\citet{mehnaz2022your} proposed two new attacks: a confidence score based model inversion attribute inference attack and a label-only model inversion attack. The key intuition of the confidence score based attack is that the target model’s returned prediction is more likely to be correct and the confidence score is more likely to be higher when it is queried with a record containing the original sensitive attribute value. Thus, the value that can get correct prediction and yield lowest loss is predicted to be the correct value of the sensitive attribute. For the label-only model inversion attack, the attacker can collect a subset of target samples which satisfy the following conditions: there is only a single value of the sensitive attribute to make the prediction correct. Once this subset is successfully collected, the attacker can train an attack model to predict the value of the sensitive attribute, assuming that the collected set is always correct. They also extend the attacks to the scenario where some of the other (non-sensitive) attributes of a target record are unknown to the adversary. Moreover, they empirically demonstrate the disparate vulnerability of model inversion attacks, i.e., specific groups in the training dataset (grouped by gender, race, etc.) could be more vulnerable to model inversion attacks. ~\citet{dibbo2023model} refined the above method by proposing a new strategy to collect training data for the attack model. Instead of using real-world data, they proposed using random generated instances and changing the value of the sensitive attribute to query the target model. Moreover, for each random instance, it is only accepted when each different sensitive attribute results in a different predicted label. Once this random instance set is gathered, an attack model is trained to predict the sensitive attribute for instances that the attacker is interested in.

\subsection{Model Extraction Attacks}

In model extraction attacks~\cite{oliynyk2023know,gong2020model}, an attacker can establish a substitute model with almost the same functionalities as the target victim model by querying the victim model. Although these attacks are not a direct threat to client data privacy in FL, they pose a threat against the server by stealing the parameters, hyper parameters, architecture, or functionality of the model~\cite{tramer2016stealing,wang2018stealing,orekondy2019knockoff}. As such, model extraction attacks serve as a privacy threat for the intellectual property (i.e., the trained model) of a company.
This is a practical impediment to the widespread deployment of FL as the trained model of the server is considered a valuable asset. Hence, threats that allow an adversary to steal that asset should be defended with a high degree of certainty. 

Model extraction attacks can also serve as a springboard for attacks that violate client privacy. For example, the adversary can use the proxy model created using model extraction to launch attacks that translate well to the original model. Due to the transferability property of such attacks (the attacks against the proxy model are accurate against the original model)~\cite{wang2021enhancing}, the adversary can use a membership inference attack to determine which client data items were used in creating the model. This is worrisome, as it has been recently shown that the best previous defense of perturbing the results of the query by the adversary as it is trying to construct the proxy model is still vulnerable to model extraction~\cite{chen2023d}. The basic idea is that this attack module, which can be integrated with any model extraction attack, can determine the difference in distribution between the unperturbed and perturbed query results and "reset" that difference.

\section{Privacy Defenses}
\label{sec:defenses}

This section discusses the core privacy defenses in FL, which are summarized in Table~\ref{tab:defense_summary} below.

\begin{table*}[ht] 
\vspace*{-2mm}
\centering
\scriptsize  
\caption{\label{tab:defense_summary} Summary of defenses against privacy attacks in federated learning}
\resizebox{\linewidth}{!}{%
\begin{tabular}{p{2.4cm} p{4.2cm} p{2.2cm} p{3.4cm} p{4.2cm}}
\toprule
\textbf{Defense class} & \textbf{Summary} & \textbf{Citations} & \textbf{Effective against} & \textbf{Resources} \\
\midrule

\textbf{Differential privacy}
& \begin{tabular}[c]{@{}l@{}}Prevents privacy leakage of any\\ individual training sample by adding\\ noise or otherwise perturbing\\ model updates.\end{tabular}
& \cite{dwork2008differential,abadi2016deep,dong2019gaussian,koskela2020computing,steinke2023privacy}
& \begin{tabular}[c]{@{}l@{}}Data reconstruction, membership\\ inference (suitable for cross-silo\\ and cross-device)\end{tabular}
& \begin{tabular}[c]{@{}l@{}}Clipping and noise added based\\ on a privacy budget. Noise injection\\ can be done at each client or at\\ the aggregated-update level.\end{tabular} \\
\midrule

\textbf{Secure aggregation}
& \begin{tabular}[c]{@{}l@{}}Individual updates are encrypted so\\ the server only sees an \emph{aggregate}\\ update, never a single client’s\\ unencrypted data.\end{tabular}
& \cite{bonawitz2017practical,secagg_bell2020secure,secagg_kadhe2020fastsecagg,zhao2021information,so2021lightsecagg}
& \begin{tabular}[c]{@{}l@{}}Data reconstruction\\ (suitable mainly for\\ cross-device)\end{tabular}
& \begin{tabular}[c]{@{}l@{}}Requires additional client-to-client\\ communication for encryption masks.\\ Typically lower overhead than\\ homomorphic encryption.\end{tabular} \\
\midrule

\textbf{Homomorphic encryption}
& \begin{tabular}[c]{@{}l@{}}Clients send encrypted updates that\\ the server can partially operate on\\ without decrypting. The server then\\ returns an encrypted aggregate to\\ clients, never seeing raw updates.\end{tabular}
& \cite{cheon2017homomorphic,zhang2020batchcrypt,fang2021privacy}
& \begin{tabular}[c]{@{}l@{}}Data reconstruction, membership\\ inference, property inference\\ (suitable for cross-silo)\end{tabular}
& \begin{tabular}[c]{@{}l@{}}High communication and\\ computation overhead for\\ encryption/decryption. Generally\\ more practical for cross-silo FL.\end{tabular} \\
\bottomrule
\end{tabular}
} 
\vspace*{-5mm}
\end{table*}

\subsection{Differential Privacy}


Differential Privacy (DP)~\cite{dwork2008differential,dwork2006calibrating} is a widely used privacy-preserving technique.  DP based defense techniques, such as DP-SGD~\cite{abadi2016deep}, add noise to the training process. Each iteration of DP-SGD satisfies a particular $(\epsilon,\delta)$-DP guarantee through the sub-sampled Gaussian Mechanism - a composition of data sub-sampling and Gaussian noise addition. Since DP is immune to post-processing, we can compose this guarantee over multiple updates to reach a final $(\epsilon,\delta)$-DP  guarantee. However, a naive composition—by summing the $\epsilon$’s from each iteration would give a huge $\epsilon$ for accurate neural networks. This usually yields a trivial but not meaningful privacy guarantee. As a result, many works have proposed more sophisticated methods for analyzing the composition of DP-SGD iterations, which can prove much tighter values of $\epsilon \leq 10 $ for the same algorithm~\cite{dong2019gaussian,koskela2020computing,mironov2017renyi}.
In addition, another line of work for differential privacy based privacy auditing is also crucial to validate that the privacy guarantee is correctly deployed. The most naive method to check if privacy guarantee is enforced requires significant computational resource because even for a single instance the training needs to be reproduced multiple times in order to get enough observation to verify the privacy guarantee. ~\citet{nasr2023tight} proposed an auditing method which requires only two runs of model training and the most recent work from ~\citet{steinke2023privacy} proposed one method which only requires a single run. Existing libraries such as Tensorflow privacy and Tensorflow federated~\cite{tensorflowprivacy} already provide implementations of many differential privacy based algorithms. However, practitioners should still be cautious about existing implementations and perform privacy auditing when necessary.


\textbf{Limitations.} DP based methods can provide a theoretical upper-bound on the effectiveness of any MI attack against any instance. However, achieving a meaningful theoretical guarantee (e.g., with a resulting $\epsilon < 5$) requires the usage of very large noises and thus results in a significant drop in accuracy. \cite{abadi2016deep} gives an example for CIFAR-10 using a two convolutional layer, two linear layer network. Without differential privacy, the network can achieve 86\% test accuracy. However, when $\epsilon=2,4,8$ are applied, the testing accuracy drops to 67\%, 70\%, and 73\% respectively. Larger $\epsilon$ values (less noise) result in a smaller accuracy drop and have still been shown to provide some empirical defense against MI attacks~\cite{li2021membership}. However, we once again highlight that such large values give up on the meaningful theoretical guarantees of DP-SGD.

DP-SGD also incurs additional time and space overhead. For example, in order to perform gradient clipping, one has to store per-sample gradients. Recent efforts to improve the privacy-utility tradeoff in DP-SGD, e.g.,~\cite{de2022unlocking,sander2023TAN}, introduce additional computational overhead.



\subsection{Secure Aggregation}

Secure aggregation (SA)~\cite{bonawitz2017practical} enables the server to aggregate local model updates from a number of users, without observing any of their model updates in the clear. Specifically, SA protocols ensure that (1) the server and any set of users do not learn any information about the local dataset of any user from the encrypted model updates in the information theoretic sense; (2) the server only learns the aggregated model; (3) correct decoding of the aggregated model in the presence of user dropout. 

\begin{figure}[]
   \vspace*{-2mm}
    \centering
    \includegraphics[scale=0.3]{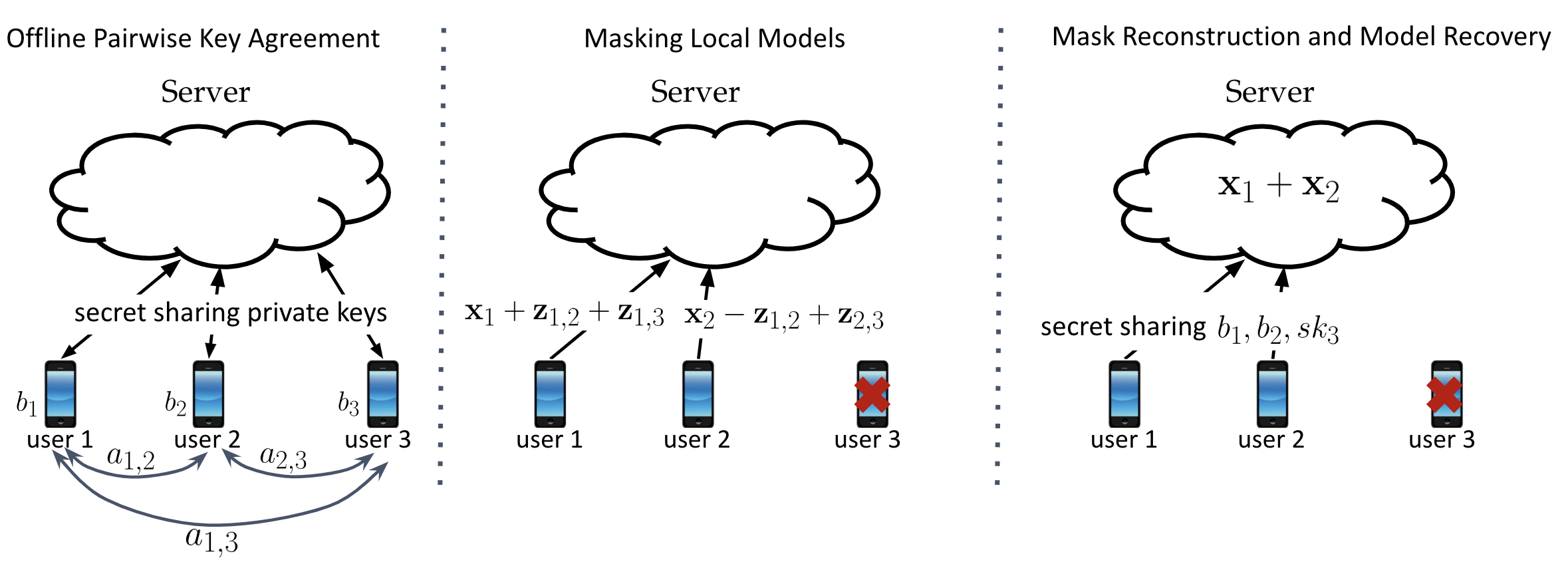}
    \vspace*{-3mm}
    \caption{An illustration of SecAgg in the example of 3 users.} 
    \label{Fig-SA}
\vspace{-5mm}
\end{figure}

The state-of-the-art for secure aggregation protocols in FL is to use additive masking to protect the privacy of individual models \cite{bonawitz2017practical,secagg_bell2020secure,secagg_so2021securing,secagg_kadhe2020fastsecagg,zhao2021information,so2021lightsecagg,so2021turbo}. SecAgg \cite{bonawitz2017practical} is the first practical protocol proposed by Google for FL that is resilient for both user failures or dropouts and collusion between users as well as the server. Particularly,  SecAgg, as depicted in Figure \ref{Fig-SA}, leverages pairwise random seed $a_{i,j}$, for $i, j \in [N]$, to be generated between each pair of users for masking the model updates. Each pair of users generate the  pairwise random seed by using a key agreement protocol (e.g., Diffie-Hellman~\cite{diffie1976new}), such that  the random seed  $a_{i,j}$ is a function of the public key $pk_i$ of user $i$   and the private key $sk_j$ of user $j$.  In addition, user $i$ creates a private random seed $b_i$ to prevent the privacy breaches that may occur if user $i$ is only delayed rather than dropped, in which case the pairwise masks alone are not sufficient for privacy protection.  For handling  users failures or dropouts, the private key and the private random seed of each user are secret shared among all other users, and can be reconstructed by the server if any user drops during the protocol, which allows for a more resilient recovery protocol against user dropouts.   User $i \in [N]$ then masks its model $\mathbf{x}_i$ as $\mathbf{\tilde{x}}_i = \mathbf{x}_i + \text{PRG}(b_i) + \sum_{j:i<j}\text{PRG}(a_{i,j}) - \sum_{j:i>j}\text{PRG}(a_{j,i})$, where PRG is a pseudo random generator, and sends it to the server. Finally, to remove the masks that involves the dropped users,  the server asks  the set of survived users  for the shares of the private keys of the dropped users and the shares of the private seeds of the survived users.  Figure  \ref{Fig-SA}  gives an example for applying  SA in the  FL setting of three users.  The cost of constructing and sharing  the masks, scales with respect to $O(N^2)$ with $N$ corresponding to the number of users, which takes the majority of execution time.

\textbf{Scalability.} As the number of clients increases, defenses such as differential privacy or homomorphic encryption typically do not have a corresponding increase in the computational or communication overheads added (they are relatively fixed regardless of the number of clients). However, SA in particular has an increase in the communication costs as the number of clients in FL increase. This largely comes from the necessity of clients to communicate among themselves to ensure that the encryption works as intended (aggregation of SA encryption masks will cancel out).

One active research direction for secure aggregation  in FL has been to   reduce the complexity of SecAgg. In particular, SecAgg+ \cite{secagg_bell2020secure} and TruboAggregate  \cite{so2020turbo} managed to reduce the quadratic complexity   of SecAgg to $O(N\log N)$. TruboAggregate leverages both  sequential training over groups of rings and lagrange coded computing \cite{yu2019lagrange}, while SecAgg+  leverages    a sparse random graph  where each user jointly encodes its model update with only  a subset of user. However, the cost of   mask reconstructions in SecAgg+ still increases as more users drop, while  TruboAggregate results in increasing  the  round/communication complexity.  There have been other secure aggregation protocols  proposed to
reduce the computation/communication  complexity of SecAgg \cite{secagg_kadhe2020fastsecagg,zhao2021information,jahani2022swiftagg}.

\textbf{Additional directions.} Another interesting research direction of SA in FL has been to further analyze its formal privacy guarantees    either over    multi-rounds  \cite{so2023securing} or from the aggregated model \cite{elkordy2022much}. In particular,  while secure aggregation protocols have provable privacy guarantees at any single round, in the sense that no information is leaked beyond the aggregate model at each round, the privacy guarantees do not extend to attacks that span multiple training rounds. The authors in \cite{so2023securing} have shown that  the  individual model may be reconstructed by leveraging  the  participation information  and the  aggregate models across multiple rounds. 

\subsection{Homomorphic Encryption}
Homomorphic encryption (HE) can be considered an extension to SecAgg within FL. While SecAgg can still expose the aggregate model to the server, HE completely shields the model information from non-data owners. HE is a cryptographic technique that enables computations to be executed on encrypted data without requiring prior decryption. In essence, HE facilitates computation while preserving data in its encrypted state. This feature is essential for upholding privacy and security in contexts necessitating processing or analyzing sensitive information. Popular schemes such as CKKS allow encrypted arithmetic on approximate numbers~\cite{cheon2017homomorphic} and are therefore interesting as a solution for encrypted machine learning. With HE, computations can be conducted on encrypted data, and subsequently, the results remain interpretable upon decryption without compromising the confidentiality of the underlying data.

Clients can employ HE to encrypt their model updates within the FL paradigm before transmission. Consequently, the server exclusively receives encrypted updates and is precluded from accessing any raw model updates. Utilizing an HE protocol, the server aggregates only encrypted weights, preserving privacy. Following aggregation, the still-encrypted updated global model is sent back to the clients. Subsequently, the clients decrypt the model using their keys for further local training. HE serves to conceal each client's contributions, thereby removing the server's access to sensitive information. Despite the computational overhead associated with HE, its implementation can markedly enhance patient data security within collaborative learning environments.

Although HE mitigates risks such as model inversion or data leakage attributable to compromised servers, it is important to consider that the final models themselves may still retain privacy-sensitive information (see membership inference in Section~\ref{sec:attacks_mem_inf}). Consequently, integrating additional privacy safeguards, such as differential privacy or partial model sharing, should be considered to prevent the potential for memorization of individual training data.

Several FL frameworks implement HE-based solutions for secure federated aggregation~\citep{jin2023fedml,cremonesi2023fed,roth2022nvidia}, thereby significantly improving privacy concerns in FL. These frameworks focus on typical applications of horizontal FL and deep learning models. Other approaches include HE in vertical FL applications, such as SecureBoost~\citep{cheng2021secureboost}, which aims for financial applications of XGBoost~\citep{Chen:2016:XST:2939672.2939785} and encrypted Kaplan-Meier for survival analysis in oncology and genome-wide association studies in FL settings~\citep{froelicher2021truly}.

Although HE holds great promise for enhancing the security of FL applications, its scalability is constrained due to larger ciphertext messages, which could be impractical in scenarios with limited bandwidth and computing resources. Consequently, HE has been predominantly investigated in the context of cross-silo or enterprise FL applications with fewer clients, which is particularly prevalent in industries such as healthcare and finance. Furthermore, HE offers additional potential for securing client training operations through its capability for training on encrypted data~\cite{lee2022privacy}.

\subsection{Discussions}

\textbf{Measurements for success.} Unlike attacks, it is quite difficult to directly compare the different privacy-preserving defenses in FL with each other. For example, while all data reconstruction attacks ultimately aim to reconstruct or recover private user data, the goals of each defense in FL are inherently different. For example, secure aggregation only aims at preventing participants (including the server) from seeing individual updates while providing no guarantees beyond this. Differential privacy, on the other hand, does not encrypt individual updates but instead provides theoretical guarantees on the effectiveness of data reconstruction or membership inference attacks. In this way, there is no universal metric that can be used to directly compare the different defenses with each other. Instead, the defenses should simply be evaluated on what they are claiming to defend against(e.g., a secure aggregation method would be a failure if an attacker could simply unencrypt individual updates). When choosing different defense methods for deployment, designers should also keep in mind the differing objectives of each defense.

Maintaining success of defenses in real-world deployment is difficult. Success in deployment would be the case where user data privacy is never breached, as any breach of privacy would be a critical problem. This is especially true given the large-scale nature of cross-device FL or the extremely sensitive nature of cross-silo FL deployments (i.e., hospital or financial data). Given how crucial it is to prevent privacy problems, it is often better to rely on certifiable defenses. An example would be differential privacy, which is currently being used for Google Gboard~\cite{xu2023federated}.

\textbf{Powerful attackers.} Recent research has shown that more powerful privacy attacks are possible in FL. In this case, individual defenses may not be enough. For example, the guarantee of secure aggregation is not enough against an attacker who can scale regardless of the number of clients~\cite{zhao2023loki} and differential privacy alone is not enough to stop an attacker from learning a representation of the data~\cite{hitaj2017deep}. Given knowledge of adversary capabilities, it could even be possible to decrease the efficiency of certain data reconstruction attacks by exploiting data heterogeneity~\cite{zhao2023loki} (current defenses work regardless of what is happening with client data heterogeneity). A more effective method against these attacks could be the use of both secure aggregation and differential privacy. However, even this combination may not be enough. If malicious clients can form a large majority of participating clients, it is still possible to single out individual updates and leak information~\cite{boenisch2023reconstructing}. Preventing these additional attacks will require more careful client selection or even stronger differential privacy guarantees, which can often come at the cost of model performance.

\section{Looking Ahead: Important Open Questions and Solution Directions}
\label{sec:look-ahead}

Compared to traditional machine learning, Federated Learning (FL) provides a unique perspective on security and privacy issues. Although learning seems to be more private due to the data being held locally on the clients, at the same time, the process still opens up new avenues for attacks due to the sharing of machine learning updates. Given inherent restrictions of FL, such as client computation and communication capabilities or other drawbacks such as in model performance when using certain defenses, research will need to continue in privacy-preserving FL. As a privacy-preserving technology by design, it is essential for FL to truly preserve the privacy of user data, even more than traditional (i.e., centralized) machine learning. At the same time, given the importance of model accuracy and performance in practice, it is also essential for FL to balance the trade-off between model utility and privacy.



The topic of Federated Learning privacy is attracting a lot of interest in research, development, and deployment. We expect that for the foreseeable future there will be robust activity on this topic with the balance between research, development, and deployment shifting over time. Here are six high-level directions that will need to be pursued and answered for FL to become truly useful in practice. 
\vspace*{-1mm}
\begin{enumerate}

\item {\em Tradeoff between utility and privacy.}
Most privacy-preserving techniques in FL fall in the tradeoff space between utility of the model (how accurate is the model under benign circumstances) and the privacy afforded to the data. A prototypical example is perturbation of the gradients being reported by the clients. This can be done through differential privacy (DP). Existing solutions show that the loss in utility is sharp for reasonable levels of privacy protection~\cite{wei2020federated,naseri2020local,kim2021federated}. There is work to be done in making this tradeoff more gradual. Broadly, compelling solutions in this space should be able to provide an adjustable tradeoff between these two dimensions. 

\item {\em From centralization to decentralization.} 
The big open question is will privacy be helped by a move toward greater decentralization, away from the coordination of a central federated server. The direction of decentralized learning has seen some initial results indicating a promising answer~\cite{koloskova2020unified,sharma2023learn,li2022learning}. However, such decentralization also brings in the vulnerability of the learning process to malicious clients~\cite{sharma2023flair}. The question of how to ensure security to malicious clients in such learning remains open. The challenge is how to make a solution that has inherent privacy benefits also secure to misbehaving participants.

\item {\em Regulatory and ethical compliance.} 
We need to be able to certify that the privacy achieved by a given solution meets a given regulation, which applies within a jurisdiction. In line with the policy and regulation discussion in Section~\ref{sec:policy}, for FL to become widely adopted in some critical application domains, it would be essential to provide privacy proofs so that the systems can be certified. Such certification is difficult enough in traditional deterministic programs and more so for the stochastic programs that are the basis of FL systems. 
As FL systems become more prevalent, it is also imperative to address ethical concerns and ensure transparency in the process towards regulatory compliance of AI systems. This involves understanding the implications of FL on various stakeholders, including data providers, model owners, and end users. Transparency measures should be implemented to provide visibility into how FL operates, including data usage, model training, and decision-making processes. Additionally, ethical guidelines should be established to govern the responsible development and deployment of FL systems, considering factors such as fairness, bias mitigation, and accountability. Incorporating ethical considerations and transparency measures will enhance trust in FL systems and promote their responsible use in various applications. They are a step towards compliance with emerging AI regulations, such as the EU AI act~\cite{woisetschlager2024federated}.

\item {\em Verification of privacy by clients.} 
In one major line of work~\cite{zhao2023loki,fowl2022robbing,pasquini2021eluding,boenisch2021curious}, we have seen that privacy can be violated by the server by sending maliciously crafted models to the clients. This triggers the question of how malicious models can be examined and detected by the clients. The comparison can be quite a daunting task because of the relative weak computing power of clients, especially mobile and embedded clients, relative to the sizes of the models. The broad question is what the stealthiness of data exfiltration attacks implies for the capability to verify manipulated models by the clients. 

\item {\em Handling diverse data sources and nodes.}
In FL, the data sources are often diverse in various aspects, such as kinds of data (images of different resolutions, videos of different lengths, etc.) and level of noise in the data (naturally present or adversarially introduced). It is imperative for private FL to be able to work with such diverse data sources, in line with the huge development of multimodal AI models. Another dimension of diversity comes from the nodes themselves, which can differ in computational power and communication connectivity. It is an ongoing challenge to enable FL with such diverse nodes, ensuring that FL does not bias the model in favor of well-connected nodes and thus leak more information from them.

\item {\em Scalability of privacy-preserving decentralized learning.} 
A prime motivation for moving toward decentralized learning is that solutions are more scalable (to a larger number of nodes and larger models) than in traditional centralized learning. However, when one interposes privacy-preserving techniques, this scalability becomes more challenging to maintain. Consider that several privacy-enhancing techniques, such as differential privacy, secure aggregation, and secure multi-party computation, which are prime tools in private FL, can introduce significant computation or communication overheads. Thus, it is an open question how ongoing advances in efficiency of these privacy-preserving techniques can be incorporated into private FL.
\item {\em Personalized FL and nonIID data.} 
A further challenge arises from the fact that client data in FL is typically nonIID (i.e., not identically distributed across clients), which can slow or hinder convergence in standard FL approaches. Personalized FL tackles this by allowing each client to maintain a set of local parameters reflecting its unique data distribution while still benefiting from a global shared model. This technique can significantly improve performance for clients whose data differ markedly from the overall population. However, ensuring privacy under personalization remains an open research area, especially when balancing the need for local customization with the constraints of private communication and aggregation. The tension between personalization, privacy, and efficient convergence continues to drive new methods in this space.


\end{enumerate}

Each of these directions presents unique challenges that must be addressed to realize the full potential of FL while safeguarding data privacy. We look forward to a field with research advances working hand-in-hand with robust practical implementations and strong policy, all leading to widespread use of Federated Learning in security and privacy-critical applications.



\begin{acks}
This material is based in part upon work supported by: (i) {\bf S. Bagchi and J. Zhao}: Cisco Research, the DEVCOM ARL Army Research Office under Contract number W911NF-2020-221, and the National Science Foundation under Grant Numbers CNS-2333487 and CNS-2038986. 
(ii) {\bf J. Li and N. Li}: Cisco Research and the National Science Foundation under Grant Numbers CNS-2247794 and CNS-2207204. 
(iii) \textbf{S. Chaterji}: DEVCOM ARL Army Research Office under Contract number W911NF-2020-221 and NSF Grant Numbers CNS-2146449 (NSF-CAREER) and CNS-2333487 (NSF-FRONTIER).
The first author Joshua Zhao was supported through part of his PhD by the Andrews Fellowship. Any opinions, findings, and conclusions or recommendations expressed in this material are those of the authors and do not necessarily reflect the views of the sponsors.
\end{acks}

\bibliographystyle{ACM-Reference-Format}
\bibliography{sample-base, saurabh, salman, ninghui, dimitris, joshua, holger, arash, chaterji}



\end{document}